\documentclass[aps, prx, reprint, superscriptaddress, longbibliography]{revtex4-2}

\usepackage{mathtools}
\usepackage{bm}
\usepackage{hyperref}
\usepackage{placeins}  
\usepackage{xcolor}  
\usepackage{amssymb}  
\usepackage[shortcuts]{extdash}  
\usepackage{multirow}
\usepackage{makecell}
\usepackage[norefs,nocites,ignoreunlbld]{refcheck}

\usepackage{mleftright}
\mleftright  

\newcommand{\mr}[1]{\mathrm{#1}}
\newcommand{\mc}[1]{\mathcal{#1}}
\newcommand{\ms}[1]{\mathsf{#1}}
\newcommand{\mb}[1]{\mathbb{#1}}
\newcommand{\hs}[1]{\hat{\mathsf{#1}}}
\newcommand{\hI}{\hat{I}}
\newcommand{\hX}{\hat{X}}
\newcommand{\hY}{\hat{Y}}
\newcommand{\hZ}{\hat{Z}}
\newcommand{\hPi}{\hat{\Pi}}
\newcommand{\Pp}[1]{\hat{P}^{(#1)}}
\newcommand{\Qp}[1]{\hat{Q}^{(#1)}}
\newcommand{\Rp}[1]{\hat{R}^{(#1)}}
\newcommand{\pt}{{\ulcorner}}

\DeclareMathOperator{\rank}{Rank}
\DeclareMathOperator{\se}{SE}
\DeclareMathOperator{\tr}{Tr}
\DeclareMathOperator{\var}{Var}

\DeclarePairedDelimiter{\floor}{\lfloor}{\rfloor}

\DeclarePairedDelimiter{\bra}{\langle}{\rvert}
\DeclarePairedDelimiter{\ket}{\lvert}{\rangle}
\DeclarePairedDelimiter{\expect}{\langle}{\rangle}
\DeclarePairedDelimiterX{\braket}[2]{\langle}{\rangle}{#1\delimsize\vert\mathopen{}#2}
\DeclarePairedDelimiterX{\ketbra}[2]{\lvert}{\rvert}{#1\delimsize\rangle\!\delimsize\langle#2}

\newcommand{\overbar}[1]{\mkern 1.5mu\overline{\mkern-1.5mu#1\mkern-1.5mu}\mkern 1.5mu}

\begin{document}


\title{Efficient tomography of microwave photonic cluster states}

\author{Yoshiki Sunada}
\email{ysunada@stanford.edu}
\altaffiliation[Present address: ]{Department of Applied Physics, Stanford University, Stanford, California 94305, USA}
\affiliation{Department of Applied Physics, Graduate School of Engineering, The University of Tokyo, Bunkyo-ku, Tokyo 113-8656, Japan}
\author{Shingo Kono}
\affiliation{RIKEN Center for Quantum Computing~(RQC), Wako, Saitama 351-0198, Japan}
\author{Jesper Ilves}
\affiliation{Department of Applied Physics, Graduate School of Engineering, The University of Tokyo, Bunkyo-ku, Tokyo 113-8656, Japan}
\author{Takanori Sugiyama}
\affiliation{Research Center for Advanced Science and Technology (RCAST), The University of Tokyo, Meguro-ku, Tokyo 153-8904, Japan}
\author{Yasunari Suzuki}
\affiliation{NTT Computer and Data Science Laboratories, Musashino, Tokyo 180-8585, Japan}
\affiliation{PRESTO, Japan Science and Technology Agency, Kawaguchi, Saitama 332-0012, Japan}
\author{Tsuyoshi Okubo}
\affiliation{Institute for Physics of Intelligence, The University of Tokyo, Bunkyo-ku, Tokyo 113-0033, Japan}
\author{Shuhei Tamate}
\affiliation{RIKEN Center for Quantum Computing~(RQC), Wako, Saitama 351-0198, Japan}
\author{Yutaka Tabuchi}
\affiliation{RIKEN Center for Quantum Computing~(RQC), Wako, Saitama 351-0198, Japan}
\author{Yasunobu Nakamura}
\affiliation{Department of Applied Physics, Graduate School of Engineering, The University of Tokyo, Bunkyo-ku, Tokyo 113-8656, Japan}
\affiliation{RIKEN Center for Quantum Computing~(RQC), Wako, Saitama 351-0198, Japan}

\date{\today}

\begin{abstract}
Entanglement among a large number of qubits is a crucial resource for many quantum algorithms.
Such many-body states have been efficiently generated by entangling a chain of itinerant photonic qubits in the optical or microwave domain.
However, it has remained challenging to fully characterize the generated many-body states by experimentally reconstructing their exponentially large density matrices.
Here, we develop an efficient tomography method based on the matrix-product-operator formalism and demonstrate it on a cluster state of up to 35 microwave photonic qubits by reconstructing its $2^{35} \times 2^{35}$ density matrix.
The full characterization enables us to detect the performance degradation of our photon source which occurs only when generating a large cluster state.
This tomography method is generally applicable to various physical realizations of entangled qubits and provides an efficient benchmarking method for guiding the development of high-fidelity sources of entangled photons.

\end{abstract}

\maketitle

\section{INTRODUCTION}

Cluster states~\cite{briegel200101persistent} are a class of many-body entangled states with applications in measurement-based quantum computation~\cite{raussendorf200105oneway,briegel200901measurementbased,yao201202experimental}, quantum error correction~\cite{schlingemann200112quantum,bell201404experimental}, quantum repeaters~\cite{azuma201504allphotonic}, quantum secret sharing~\cite{markham200810graph,bell201412experimental}, and quantum metrology~\cite{friis201706flexible,shettell202003graph}.
They are also remarkable for the existence of experimental protocols for generating an arbitrarily large cluster state using a finite hardware by time-domain multiplexing~\cite{lindner200909proposal,menicucci201106temporalmode,alexander201803universal}.
Demonstrations of such protocols have generated linear cluster states of itinerant optical~\cite{yokoyama201312ultralargescale,schwartz201610deterministic,istrati202010sequential,thomas202208efficient} and microwave~\cite{besse202009realizing} photons.
These one-dimensional cluster states have been further entangled into two-dimensional cluster states, which are a universal resource for measurement-based quantum computation~\cite{asavanant201910generation,larsen201910deterministic,ferreira202405deterministic,osullivan202409deterministic}.

To fully characterize an experimentally-generated cluster state, its many-body density matrix needs to be reconstructed from a complete set of measurements.
The reconstructed density matrix can then be used to calculate such metrics as the localizable entanglement~\cite{verstraete200401entanglement,popp200504localizable}, which can evaluate the usefulness of an entangled state for measurement-based quantum computation~\cite{vandennest200610universal}.
However, reconstructing the density matrix of a large entangled state has remained challenging because the conventional quantum state tomography becomes exponentially costly as the number of qubits increases.
To circumvent this difficulty, large-scale cluster states have been characterized in one of two ways: evaluating an entanglement witness~\cite{yokoyama201312ultralargescale,istrati202010sequential,thomas202208efficient} or performing a quantum process tomography of each operation in the entanglement generation protocol~\cite{schwartz201610deterministic,besse202009realizing}.
In the first approach, theoretical work is required to construct a witness operator for each purpose, such as certifying genuine multipartite entanglement~\cite{vanloock200305detecting,toth200502detecting} and lower-bounding the quantum state fidelity~\cite{thomas202208efficient} or localizable entanglement~\cite{nutz201706proposal}.
The second approach evaluates the individual steps of the protocol in isolation, which means that a numerical extrapolation is required to evaluate the entire protocol.
This approach also requires the assumption that the photon emission process executed during the protocol is identical to when it is executed in isolation (Table~\ref{tab:comparison}).
Recently, Ref.~\citenum{osullivan202409deterministic} demonstrated a tomography of a 20-qubit quasi-two-dimensional microwave photonic cluster state by implementing an iterative maximum-likelihood algorithm using a matrix-product-operator~(MPO) representation of the density matrix~\cite{baumgratz201312scalable}.
However, for a large problem, the convergence of the iterative maximum-likelihood algorithm is significantly slower than that of a direct optimization algorithm~\cite{moroder201210permutationally}.
Moreover, the convergence cannot be guaranteed without further reducing the convergence rate~\cite{rehacek200704diluted}.

\begin{table*}
\caption{Comparison of tomography methods for entangled photonic qubits.}
\label{tab:comparison}
\begin{ruledtabular}
\begin{tabular}{lccc}
& Conventional tomography & \makecell{Process tomography and\\numerical extrapolation\\ \cite{schwartz201610deterministic,besse202009realizing}} & MPO reconstruction \\
\hline
Number of parameters to estimate & $4^N-1$ & $O(1)$ & $O(N)$ \\
Assumes sequentially emitted photons? & No & Yes & Yes \\
Assumes identical emission process? & No & Yes & No \\
\end{tabular}
\end{ruledtabular}
\end{table*}

\begin{table*}
\caption{Comparison of tomography methods based on MPO reconstruction.}
\label{tab:comparison2}
\begin{ruledtabular}
\begin{tabular}{lccc}
& \makecell{Direct inversion\\ \cite{baumgratz201307scalable}} & \makecell{Iterative maximum likelihood\\ \cite{baumgratz201312scalable,osullivan202409deterministic}} & This work \\
\hline
Requires iterative optimization of MPO? & No & Yes (slow) & Yes (fast) \\
Robust against statistical noise of high-order correlations? & No & No & Yes \\
Guarantees positivity of reconstructed MPO? & No & Yes & No \\
\end{tabular}
\end{ruledtabular}
\end{table*}

Here, we propose and experimentally demonstrate an efficient tomography method for reconstructing the many-body density matrix of a sequentially emitted chain of entangled photonic qubits.
This is made possible by the fact that the exponentially large density matrix can be parameterized efficiently using an MPO representation even if there are coherent and incoherent errors in the photon emission process.
To demonstrate the tomography method, we use a superconducting qubit to generate linear cluster states of up to 35 microwave photonic qubits in the discrete-variable basis.
We then use a quantum-limited amplifier to measure the quadrature observables of the photonic qubits and efficiently obtain the complete set of correlations among five consecutive qubits.
An MPO is then fitted to the measured correlations using the Gauss--Newton weighted least-squares algorithm, which is known to converge linearly or better given a good initial guess.
The weighted fitting makes our method robust against the large statistical noise inherent in high-order correlation measurements and enables us to propagate the measurement uncertainties to the parameters of the MPO.
To obtain the good initial guess, we derive an approximate reconstruction formula for the MPO in terms of the measured correlations by reformulating the direct inversion procedure introduced in Ref.~\citenum{baumgratz201307scalable}.
This reformulation also provides a method to directly estimate the required bond dimension of the MPO representation from the measured correlations.
These features make our tomography method an attractive alternative to methods based on direct inversion~\cite{baumgratz201307scalable} or the iterative maximum-likelihood algorithm~\cite{baumgratz201312scalable,osullivan202409deterministic}.
Table~\ref{tab:comparison2} summarizes the differences between these tomography methods.

Using the reconstructed MPO representations of the cluster states, we calculate the quantum state fidelity and localizable entanglement.
The localizable entanglement is found to be significantly larger than the lower bound determined by stabilizer measurements~\cite{nutz201706proposal}.
We demonstrate that the entanglement in the 10-qubit cluster state persists for up to a length of seven consecutive photons.
We also find that this length decreases for the larger cluster states, which contradicts the numerical extrapolation of the results for the smaller cluster states.
Our observation suggests that the successful generation of a smaller cluster state does not guarantee the generation of a larger cluster state, highlighting the importance of efficient tomography methods for large entangled states.

\section{EFFICIENT TOMOGRAPHY}

In this section, we first show that the process tensor representing the sequential emission of photonic qubits can be transformed into an MPO.
This means that the many-body density matrix of the emitted photonic qubits can be efficiently represented using a linear number of parameters with respect to the number of qubits.
However, a parameter in this representation does not directly correspond to a measurable quantity of the photon chain.
To realize an efficient tomography, one needs to choose a small set of measurable quantities to which these parameters can be fitted.
In the second part of this section, we describe how a linear cluster state can be tomographically reconstructed from a set of local correlations, which can be measured in parallel using a constant number of measurement settings regardless of the number of qubits.

\subsection{Efficient parameterization of sequentially generated entanglement}

In general, an $N$-qubit density matrix $\hat{\rho}$ can be represented in the Pauli basis as
\begin{equation}
    \hat{\rho} = \frac{1}{2^N} \sum_{i_1, \ldots, i_N \in \{0, 1, 2, 3\}}
        \rho_{i_1, \ldots, i_N} \Pp{i_1}_1 \cdots \Pp{i_N}_N,
\end{equation}
where $\rho_{i_1, \ldots, i_N}$ is a $4 \times \cdots \times 4$ real tensor, and $\Pp{0}_s = \hI_s$, $\Pp{1}_s = \hX_s$, $\Pp{2}_s = \hY_s$, and $\Pp{3}_s = \hZ_s$ are the Pauli operators on the $s$th qubit for $s = 1, \ldots, N$.
In the conventional quantum state tomography, each parameter $\rho_{i_1, \ldots, i_N}$ is estimated from an $N$-qubit correlation measurement as $\rho_{i_1, \ldots, i_N} = \expect{\Pp{i_1}_1 \cdots \Pp{i_N}_N}$.
Taking into account the unit-trace constraint, $\rho_{0, \ldots, 0} = \tr[\hat{\rho}] = 1$, the number of free parameters is $4^N - 1$, which makes this procedure exponentially difficult as the number of qubits increases.

In the MPO representation, the $4 \times \cdots \times 4$ tensor $\rho_{i_1, \ldots, i_N}$ is decomposed into a contraction of $N$ tensors: a $4 \times D$ tensor $\ms{A}_1$, $D \times 4 \times D$ tensors $\ms{A}_2, \ldots, \ms{A}_{N-1}$, and a $D \times 4$ tensor $\ms{A}_N$~\cite{cirac202112matrix}.
The contraction can be expressed as matrix multiplications by slicing each tensor along the four-dimensional axis:
\begin{equation} \label{eq:mpo}
    \rho_{i_1, \ldots, i_N} = A^{(i_1)}_1 A^{(i_2)}_2 \cdots A^{(i_{N-1})}_{N-1} A^{(i_N)}_N.
\end{equation}
Here, $A^{(i_1)}_1$ is a $D$-dimensional row vector, $A^{(i_2)}_2, \ldots, A^{(i_{N-1})}_{N-1}$ are $D \times D$ matrices, and $A^{(i_N)}_N$ is a $D$-dimensional column vector.
The dimension $D$ of the axes being contracted is called the bond dimension of the MPO.
The bond dimension $D$ determines the expressivity of the MPO and is closely related to the entanglement entropy across the bond~\cite{zanardi200103entanglement,prosen200709operator}.
Note that the decomposition is not unique and is possible with any value of $D$ sufficiently large, although a smaller $D$ gives a more efficient MPO representation.
Appendix~\ref{app:mpo-formalism} describes how to efficiently calculate multi-qubit correlations, quantum state fidelities, and effects of local errors using the MPO representation.

The MPO representation has been used to efficiently parameterize and characterize a quantum state generated by a system with local interactions such as a linear ion trap~\cite{baumgratz201307scalable,lanyon201712efficient}.
Here, we show that a sequentially-emitted chain of entangled photonic qubits can also be efficiently represented by an MPO because of the locality in the time dimension.
We expand on Ref.~\citenum{schon200509sequential}, which showed that a chain of photonic qubits generated by a $d$-level system~(qudit) can be represented by a matrix product state~(MPS) with a bond dimension of at most $d$ if the qudit--photon system remains in a pure state.
We use the concepts of mixed-state quantum circuits~\cite{aharonov1998quantum} and tensor networks~\cite{biamonte202001lectures} to confirm the suggestion in Ref.~\citenum{schon200509sequential} that an MPO can be used to allow for mixed states and decoherence.

\begin{figure}
\centering
\includegraphics{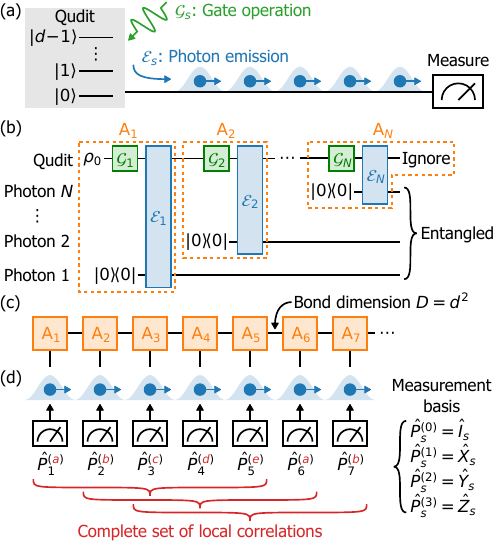}
\caption{Efficient tomography of sequentially-emitted entangled photonic qubits.
(a)~Sequential emission of entangled photonic qubits by a $d$-level system~(qudit).
(b)~Mixed-state quantum-circuit representation, which can be interpreted as a tensor network corresponding to the density matrix of the emitted photonic qubits.
(c)~Matrix product operator~(MPO) obtained by contracting the tensors in groups.
This is an efficient parameterization of the many-body density matrix of the generated photonic qubits because the number of parameters only increases proportionally with the number of photonic qubits, $N$.
(d)~Efficient choice of measurement bases which makes the number of measurement settings independent of $N$.
Here, we measure the complete set of local correlations of up to five consecutive photonic qubits, which is sufficient for reconstructing the density matrix of a linear cluster state.
}
\label{fig:figure1}
\end{figure}

Figure~\ref{fig:figure1}(a) depicts a qudit sequentially emitting a chain of photonic qubits.
This process can be modeled using the mixed-state quantum circuit shown in Fig.~\ref{fig:figure1}(b).
The initial density matrix of the qudit is represented by a $d^2$-dimensional real vector $\rho_0$ by assuming a Hermitian matrix basis such as the generalized Gell-Mann matrices~\cite{kimura200308bloch}.
Then, the effect of a gate operation on the qudit, including any coherent and incoherent errors, is to multiply the vector by a $d^2 \times d^2$ process matrix $\mc{G}_s$.
Similarly, a conditional photon emission from the qudit is a gate operation between the qudit and the photonic qubit and is represented by a $d^2 \times 2^2 \times d^2 \times 2^2$ process tensor $\mc{E}_s$.
Note that the initial state of the photonic qubit is chosen to be the vacuum state $\ketbra{0}{0}$ without loss of generality.
A chain of $N$ photonic qubits is generated by alternating between the qudit gates $\mc{G}_s$ and photon emissions $\mc{E}_s$, which may differ each time as a part of the experimental protocol or because of its imperfect execution.
Finally, an entangled state of the $N$ photonic qubits is obtained by ignoring the qudit state at the end of the protocol, which is equivalent to taking the partial trace over the qudit.

This quantum circuit can be interpreted as a tensor-network diagram corresponding to the density matrix of the emitted photonic qubits.
In a tensor-network diagram, a node with $n$ edges represents a tensor with $n$ indices, an edge connecting two nodes represents a tensor contraction, and a dangling edge represents an index of the resulting tensor.
Then, the tensors can be contracted in groups as shown by the dashed orange boxes in Fig.~\ref{fig:figure1}(b) to obtain a tensor network shown in Fig.~\ref{fig:figure1}(c), which corresponds to Eq.~\eqref{eq:mpo} and represents an MPO.
The dimension of the tensor contractions represented by the horizontal edges corresponds to the bond dimension of the MPO.
Therefore, the $N$-qubit density matrix of the generated photonic qubits can be efficiently parameterized using an MPO with a bond dimension of at most $D = d^2$, whose number of parameters only grows linearly with $N$.
Notably, this representation is efficient even if additional energy levels of the photon emitter are unintentionally involved, under the reasonable assumption that the number of the participating energy levels does not increase exponentially with $N$.
Thus, the MPO representation allows one to perform the quantum state tomography of a sequentially generated entanglement without estimating each of the exponentially many parameters of the density matrix.

\subsection{Efficient choice of measurement settings}

Because the MPO representation reduces the number of parameters describing the density matrix, one may expect that the parameters can be estimated from a small set of correlation measurements.
However, choosing an appropriate subset from the exponentially large set of possible $N$-qubit correlation measurements is a nontrivial problem.

Our strategy for choosing an efficient set of correlation measurements is based on Ref.~\citenum{baumgratz201307scalable}, which showed that an $N$-qubit state represented by an MPO can be reconstructed from its local reductions if the ``reconstructibility condition'' described below is satisfied.
A local reduction of a chain of qubits is defined as the state obtained by selecting $L$ consecutive qubits and ignoring the others.
Because the local reduction is an $L$-qubit state, it is determined by the complete set of $L$-qubit correlation measurements.
Taking $L = 5$ as an example, the local correlations can be denoted as
\begin{equation} \label{eq:pauli-correlation}
    C^{(a,b,c,d,e)}_s
    \coloneqq \expect{\Pp{a}_s \Pp{b}_{s+1} \Pp{c}_{s+2} \Pp{d}_{s+3} \Pp{e}_{s+4}},
\end{equation}
where $a, b, c, d, e \in \{0, 1, 2, 3\}$ and $s = 1, \ldots, N-4$.
The MPO representation of the $N$-qubit state can be reconstructed from the set of local correlations by following the procedure introduced in Ref.~\citenum{baumgratz201307scalable}.
Appendix~\ref{app:mpo-inversion} gives an explicit formula for the reconstructed MPO in terms of $C^{(a,b,c,d,e)}_s$.
The reconstructibility condition is fulfilled by the vast majority of MPOs whose bond dimension $D$ satisfies
\begin{equation} \label{eq:d-l-condition}
    D \le 4^{\floor{(L - 1) / 2}},
\end{equation}
where $\floor{\cdot}$ denotes the floor function~(see Appendix~\ref{app:mpo-inversion} for the full statement of the reconstructibility condition).
The exceptional cases form a zero-volume subset within the space of all MPOs with a bond dimension of $D$.
In these cases, the MPO is unreconstructible because one or more of the linear systems of equations which appear in the reconstruction procedure has no solution.

One might expect that a linear cluster state, which is an MPO with a bond dimension of $D = 4$, can be reconstructed from the complete set of three-qubit local correlations because $D = 4$ and $L = 3$ satisfy Eq.~\eqref{eq:d-l-condition}.
However, as we show in Appendix~\ref{app:mpo-inversion}, a linear cluster state is an exceptional case in the sense that its reconstruction requires $L = 5$.
Although decoherence and imperfections in the experiment will likely bring the state out of the set of exceptional cases, the reconstruction procedure is sensitive to statistical noise in the data if the state being measured is close to an unreconstructible state.
Therefore, for evaluating the performance of a high-fidelity source of linear cluster states, the generated state should be reconstructed from the five-qubit local correlation measurements.

The complete set of five-qubit local correlations can be efficiently measured by choosing the measurement basis for each qubit as shown in Fig.~\ref{fig:figure1}(d), where every fifth qubit is measured in the same basis.
A local correlation $C^{(a,b,c,d,e)}_s$ can be obtained from such a measurement setting by ignoring the measurement outcomes of all but five consecutive qubits.
Thus, the number of measurement settings required to reconstruct the density matrix of a sequentially generated $N$-qubit entanglement has been reduced to a constant which is independent of $N$.

\section{EXPERIMENT}

We demonstrate the efficient tomography method on linear cluster states of itinerant microwave photons generated using a superconducting qubit.
In this section, we describe how we generate a microwave photonic cluster state and measure the correlations among the photonic qubits.

\subsection{Generating a microwave photonic cluster state}

\begin{figure}
\centering
\includegraphics{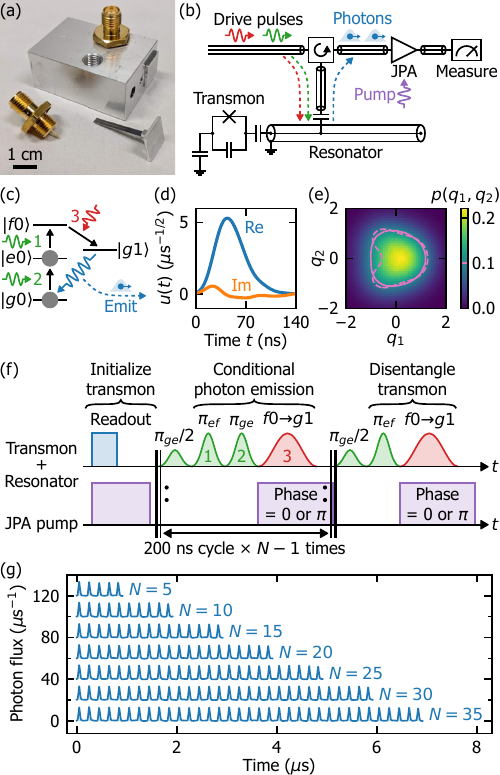}
\caption{Generation and measurement of a microwave photonic linear cluster state.
(a)~Photograph of the disassembled~(front) and assembled~(back) microwave photon sources, which consist of a superconducting transmon qubit and a microwave resonator on a silicon substrate inside an aluminum cylindrical cavity. The threaded connector with a long center pin is used to couple a coaxial cable to the resonator.
(b)~Circuit model of the photon source and a schematic of the measurement chain, which includes a Josephson parametric amplifier~(JPA) for phase-sensitive quantum-limited amplification.
(c)~Energy-level diagram of the photon source. The numbered arrows correspond to the steps comprising the conditional photon emission operation. 
(d)~Measured mode function $u(t)$ of the photonic qubit.
(e)~Probability density $p(q_1, q_2)$ of measuring the quadrature values $q_1$ and $q_2$ for the first and second photonic qubits, respectively, of the generated five-qubit cluster state.
The solid and dashed lines are the contour lines for the measured data and an ideal cluster state under the effect of the same measurement inefficiency, respectively.
(f)~Pulse sequence for generating an $N$-qubit linear cluster state.
(g)~Measured photon flux of the cluster states. The traces are offset vertically by units of 20~$\mu$s$^{-1}$ for clarity.}
\label{fig:figure2}
\end{figure}

To generate a linear cluster state, we use a protocol based on the proposal in Ref.~\citenum{lindner200909proposal}, which uses an electron confined in a semiconductor quantum dot to generate a linear cluster state of optical photons.
This protocol inspired the experimental demonstrations using a dark exciton~\cite{schwartz201610deterministic}, an InGaAs quantum dot~\cite{istrati202010sequential}, a rubidium atom~\cite{thomas202208efficient}, and superconducting qubits~\cite{besse202009realizing,ferreira202405deterministic}.
Our protocol is based on a microwave-activated photon--qudit interaction and is different from the previous works in the microwave domain in the sense that it does not use any frequency-tunable qubit or tunable coupler.
This protocol is promising for increasing the fidelity and scale of the cluster states because tuning mechanisms tend to increase the decoherence rates and wiring complexity of the photon source.

Figure~\ref{fig:figure2}(a) shows a photograph of the photon source, which consists of a fixed-frequency superconducting transmon qubit~\cite{koch200710chargeinsensitive} and a coaxial transmission line resonator~\cite{axline201607architecture}.
The resonator couples to a coaxial cable through an intrinsic Purcell filter to enable a fast photon generation while minimizing the energy relaxation of the superconducting qubit~\cite{sunada202204fast}.
The circuit model of the photon source and a schematic of the measurement chain are shown in Fig.~\ref{fig:figure2}(b) and are described in more detail in Appendix~\ref{app:sample-and-setup}.
Figure~\ref{fig:figure2}(c) shows the energy-level diagram, where $\ket{g}$, $\ket{e}$, and $\ket{f}$ are the ground, first-excited, and second-excited states of the transmon and $\ket{0}$ and $\ket{1}$ are Fock states of the resonator.
A photonic qubit is generated using the microwave-induced coupling between the $\ket{f0}$ and $\ket{g1}$ states~\cite{pechal201410microwavecontrolled,zeytinoglu201504microwaveinduced}, which has been used to generate a shaped photon~\cite{pechal201410microwavecontrolled} and a time-bin photonic qubit~\cite{kurpiers201910quantum,ilves202004ondemand}.
The $\ket{f0}$\textrightarrow$\ket{g1}$ transition converts the $\ket{f}$ population of the transmon to a single-photon excitation of the resonator, which is then emitted as an itinerant microwave photon into the coaxial cable which couples to the resonator.
Figure~\ref{fig:figure2}(d) shows the mode function $u(t)$ of the emitted photon, which is obtained by generating a photonic qubit in the superposition state $(\ket{0} + \ket{1}) / \sqrt{2}$ and measuring its averaged waveform.
The spontaneous emission from the resonator occurs simultaneously with the $\ket{f0}$\textrightarrow$\ket{g1}$ transition because the resonator has a large external linewidth of 18.2~MHz relative to the microwave-activated $\ket{f0}$--$\ket{g1}$ coupling strength.
This allows for the possibility to shape the photonic qubit into a time-symmetric envelope suitable for quantum communication by optimizing the shape of the $\ket{f0}$\textrightarrow$\ket{g1}$ drive pulse~\cite{kurpiers201806deterministic,kurpiers201910quantum}.

Figure~\ref{fig:figure2}(f) shows the pulse sequence for generating an $N$-qubit linear cluster state~(see Appendix~\ref{app:cluster-state-protocol} for more details).
The transmon is initialized to the ground state using a quantum nondemolition readout and post-selection.
Then, we repeat the 200-ns cycle consisting of a 20-ns $\pi/2$ rotation in the qubit subspace $\{\ket{g0}, \ket{e0}\}$ and a photon emission conditioned on the qubit being in $\ket{e0}$.
The conditional photon emission is composed of two 20-ns $\pi$-pulses which induce $\ket{e0}$\textrightarrow$\ket{f0}$ and $\ket{g0}$\textrightarrow$\ket{e0}$ transitions in this order, followed by a $\ket{f0}$\textrightarrow$\ket{g1}$ drive, which is a chirped raised-cosine pulse~\cite{ilves202004ondemand} with a duration of 100~ns.
After repeating the 200-ns cycle $N - 1$ times, another $\pi/2$ rotation is performed and the transmon is disentangled from the generated photons by completely emitting its excitation as a photon.
We used this pulse sequence to generate $N$-qubit cluster states for $N = 5$, 10, 15, 20, 25, 30, and 35.
Figure~\ref{fig:figure2}(g) shows the measured photon flux of the cluster states.

\subsection{Measuring the microwave photonic qubits}

Each photonic qubit generated by this protocol is defined in the $\{\ket{0}, \ket{1}\}$ subspace of a microwave pulse mode.
Such a qubit can be measured in the Pauli $\hat{Z}$ basis using a microwave photon detector~\cite{inomata201607single,kono201803quantum,besse201804singleshot,balembois202401cyclically}.
Pauli $\hat{X}$ and $\hat{Y}$ measurements can be realized by absorbing the photon into a superconducting qubit, which requires temporally modulating the coupling strength to match the mode function of the photon~\cite{kurpiers201806deterministic,kurpiers201910quantum}.
Here, we take an experimentally simpler approach of measuring the quadrature observables $\hat{q}_s \coloneqq (\hat{a}_s^\dag + \hat{a}_s) / \sqrt{2}$ and $\hat{p}_s \coloneqq i (\hat{a}_s^\dag - \hat{a}_s) / \sqrt{2}$, where $\hat{a}_s$ is the annihilation operator of the pulse mode for the $s$th photonic qubit.
The data analysis methods in this work are directly applicable to experiments in the optical domain because the same quantities can be measured using optical homodyne detection~\cite{lvovsky200903continuousvariable}.

We measure the quadrature observables of the microwave photonic qubits using a flux-driven Josephson parametric amplifier~(JPA)~\cite{kono201803quantum}.
By driving the JPA with a 140-ns pulse at twice the signal frequency, the signal can be phase-sensitively amplified.
For each qubit, either the $\hat{q}_s$ or $\hat{p}_s$ quadrature is amplified by switching the phase of the pump pulse between 0 and $\pi$.
The quadrature values are obtained by calculating the overlap integral between the amplified waveform and the mode function $u(t)$ of the photonic qubit measured with the method described in Appendix~\ref{app:mode-function}.
Figure~\ref{fig:figure2}(e) shows the two-dimensional histogram of the quadrature values measured for the first and second photonic qubits of the five-qubit cluster state.
The visible correlation between $\hat{q}_1$ and $\hat{q}_2^2$ corresponds to the Pauli-basis correlation $\expect{\hX_1 \hZ_2} = 1$, which is characteristic of a cluster state.
Whereas this histogram is not corrected for the measurement inefficiency, we correct the data used in the following analysis using the estimated measurement efficiency of $0.391 \pm 0.004$ (see Appendix~\ref{app:pauli-tomography} for details).
We also numerically correct for the phase offsets between the generated photonic qubits and the measured quadratures by maximizing the expectation values of the stabilizer operators of the cluster state~(see Appendix~\ref{app:full-tomography} for details).

To measure a five-qubit correlation, we choose for each photonic qubit either $\hat{q}_s$ or $\hat{p}_s$ as the measurement basis, perform $10^7$ measurements with a repetition period of 10~$\mu$s, and calculate the set of moments
\begin{equation} \label{eq:quadrature-correlation}
    \expect{\Qp{a}_s \Qp{b}_{s+1} \Qp{c}_{s+2} \Qp{d}_{s+3} \Qp{e}_{s+4}},
\end{equation}
where $a, b, c, d, e \in \{0, \ldots, 5\}$ and 
\begin{subequations} \begin{align}
    \Qp{0}_s & \coloneqq \hat{q}_s^0, &
    \Qp{1}_s & \coloneqq \hat{p}_s^0, \label{eq:Q-a} \\
    \Qp{2}_s & \coloneqq \hat{q}_s^1, &
    \Qp{3}_s & \coloneqq \hat{p}_s^1, \\
    \Qp{4}_s & \coloneqq \hat{q}_s^2, &
    \Qp{5}_s & \coloneqq \hat{p}_s^2. \label{eq:Q-c}
\end{align} \end{subequations}
These multivariate moments are then converted to the Pauli-basis correlations $C^{(a,b,c,d,e)}_s$ using the formula derived in Appendix~\ref{app:pauli-tomography}, which is a multi-qubit generalization of the equalities
\begin{subequations} \begin{align}
    \expect{\hX_s} & = \sqrt{2} \expect{\hat{q}_s}, \label{eq:pauli-quadrature-a} \\
    \expect{\hY_s} & = \sqrt{2} \expect{\hat{p}_s}, \\
    \expect{\hZ_s} & = 2 - \expect{\hat{q}_s^2} - \expect{\hat{p}_s^2} \label{eq:pauli-quadrature-c}
\end{align} \end{subequations}
for a pulse mode containing up to one photon.
Only $2^5 = 32$ measurement settings, i.e., the combination of the JPA phase for each photonic qubit, are required to measure the complete set of five-qubit local correlations for the entire photon chain.

\section{DATA ANALYSIS}

To reconstruct the MPO representation of the density matrix of the generated photonic qubits, one needs to find the MPO which fits the set of measured local correlations.
We begin this section by showing how the required bond dimension of the MPO can be determined from the measured local correlations.
We then describe the least-squares method which we use to fit the MPO to the data and propagate the statistical uncertainties in the data to the parameters of the MPO.

\subsection{Estimating the required bond dimension} \label{sec:estimating-d}

\begin{figure}
\centering
\includegraphics{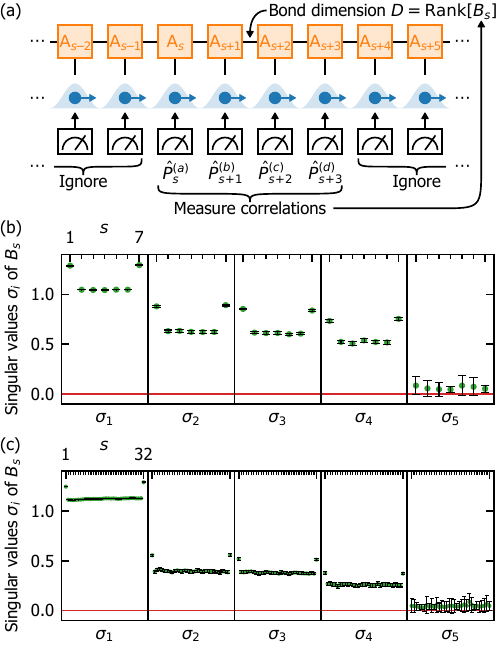}
\caption{Estimating the required bond dimension of the MPO representation using the measured local correlations.
(a)~Schematic of the procedure.
(b),~(c)~Five largest singular values $\sigma_1, \ldots, \sigma_{5}$ of the four-qubit local correlation matrices $B_s$~($s = 1, \ldots, N-3$) measured for the $N$-qubit linear cluster states with $N = 10$ and 35, respectively.
Error bars represent the standard errors of the singular values estimated by the first-order propagation of the statistical errors of the correlation data.
Because only four singular values of any $B_s$ are significantly larger than zero, the bond dimension of these cluster states can be estimated as $D = 4$.}
\label{fig:figure3}
\end{figure}

The state of a photon chain generated by our photon source has a bond dimension of at most $4^2 = 16$ because only four energy levels participate in the photon generation even if there are leakage errors into the fourth energy level of the transmon.
However, to reduce the computational cost of the MPO reconstruction, it is desirable to know in advance if the measured state has a more efficient MPO representation with a smaller bond dimension.
In fact, the state generated by our experimental protocol can be expected to have a bond dimension of $2^2 = 4$ if there is no leaked population outside the qubit subspace $\{ \ket{g0}, \ket{e0} \}$ after each conditional photon emission.

Here, we estimate the bond dimension required to express the measured state by calculating the ranks of $16 \times 16$ matrices $B_s$ constructed from the four-qubit local correlations,
\begin{equation}
    (B_s)_{4a+b, 4c+d} \coloneqq
    \expect{\Pp{a}_s \Pp{b}_{s+1} \Pp{c}_{s+2} \Pp{d}_{s+3}},
\end{equation}
as shown in Fig.~\ref{fig:figure3}(a).
As we show in Appendix~\ref{app:reducing-d}, the required bond dimension between the $(s+1)$th and $(s+2)$th photonic qubits is given by $\rank[B_s]$.
This means that the measured $N$-qubit state has an MPO representation with a bond dimension of $D$ if $\rank[B_s] \le D$ for every $s = 1, \ldots, N-3$.
Note that $B_s$ should be constructed from the local correlations of a larger number of consecutive qubits if the measured state can have a bond dimension larger than 16.
In general, the complete set of $(2l)$-qubit local correlation measurements allows one to perform this analysis for a state with a bond dimension of up to $4^l$.

Since the measured four-qubit correlations $B_s$ contain statistical noise, we estimate $\rank[B_s]$ by counting the number of singular values of $B_s$ which are significantly larger than their statistical uncertainties.
The statistical uncertainties are estimated by the first-order propagation of the uncertainties of the quadrature measurements to the correlation matrix $B_s$ then to the singular values~(see Appendix~\ref{app:reducing-d} for details).
Figures~\ref{fig:figure3}(b) and~(c) show the five largest singular values $\sigma_1, \ldots, \sigma_{5}$ of the four-qubit correlation matrices $B_s$ measured for an $N$-qubit linear cluster state with $N = 10$ and 35, respectively.
Because only four singular values of any $B_s$ are significantly larger than zero, the bond dimension required to express these cluster states can be estimated as $D = 4$, up to the statistical uncertainties of the quadrature measurements.
Note that this is only a qualitative assessment and that a statistical test for the rank of a noisy matrix should be developed in the future to rigorously determine the bond dimension of the measured state.

\subsection{Least-squares fitting by an MPO} \label{sec:mpo-fitting}

If the exact values of the five-qubit correlations were known, one would be able to reconstruct the MPO representations of the cluster states from the correlation matrices using the explicit formula given in Appendix~\ref{app:mpo-inversion}.
However, this procedure is sensitive to any noise in the correlation measurements because it involves computing the Moore--Penrose pseudoinverses of ill-conditioned matrices~\cite{baumgratz201307scalable}.
In this work, we instead perform a least-squares fitting of the MPO to the measured correlations.
The fit parameters are the elements of the matrices $A^{(i_s)}_s$ comprising the MPO, and the fit residuals to be minimized are the differences between the measured correlations $C^{(a,b,c,d,e)}_s$ and the corresponding correlations calculated from the MPO as
\begin{equation} \label{eq:mpo-correlation}
    A^{(0)}_1 \cdots A^{(0)}_{s-1}
    A^{(a)}_s A^{(b)}_{s+1} A^{(c)}_{s+2} A^{(d)}_{s+3} A^{(e)}_{s+4}
    A^{(0)}_{s+5} \cdots A^{(0)}_N.
\end{equation}
To impose the unit-trace constraint of a density operator and to reduce the number of fit parameters by approximately half, the MPO is restricted to the ``standard form'' introduced in Appendix~\ref{app:mpo-standard-form}.
We use the analytic formula for the partial derivative of Eq.~\eqref{eq:mpo-correlation} with respect to each fit parameter to implement the Gauss--Newton algorithm, which can efficiently solve a nonlinear least-squares problem.
The partial derivatives also allow us to propagate the statistical uncertainties of the measured correlations $C^{(a,b,c,d,e)}_s$ to the uncertainties of the reconstructed MPO.
To provide a good initial guess to the fitting algorithm, we follow the procedures in Appendices~\ref{app:mpo-inversion} and \ref{app:reducing-d} to reconstruct an approximate MPO by compressing the bond dimension using truncated singular value decompositions.

In this work, the measured Pauli-basis correlations $C^{(a,b,c,d,e)}_s$ contain statistical uncertainties of various magnitudes because they are calculated using Eqs.~\eqref{eq:pauli-quadrature-a}\==\eqref{eq:pauli-quadrature-c} from the quadrature moments.
In particular, Pauli-basis correlations which measure many of the qubits in the $\hZ_s$ basis have large uncertainties because they are calculated from high-order quadrature moments.
To avoid overfitting to the large statistical errors in such correlation measurements, we perform a weighted least-squares fitting, where each residual is weighted by the reciprocal of the standard error of the fit target~(see Appendix~\ref{app:10-qubit} for the fit result for the 10-qubit cluster state).

\subsection{Results of MPO reconstruction}

\begin{figure*}
\centering
\includegraphics{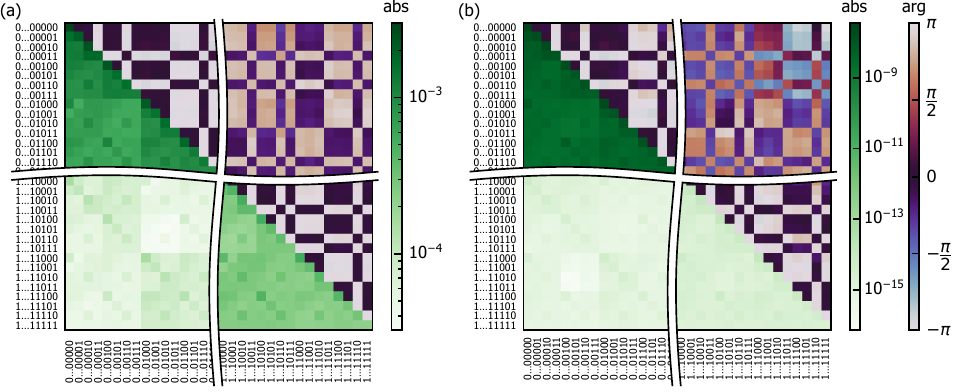}
\caption{Corners of the reconstructed density matrices of the (a)~10-qubit and (b)~35-qubit cluster states. Absolute values are plotted in the diagonal and the lower-left triangle, and the complex arguments in the upper-right triangle. See Appendix~\ref{app:10-qubit} for the full density matrix of the 10-qubit cluster state.}
\label{fig:figure4}
\end{figure*}

\begin{figure}
\centering
\includegraphics{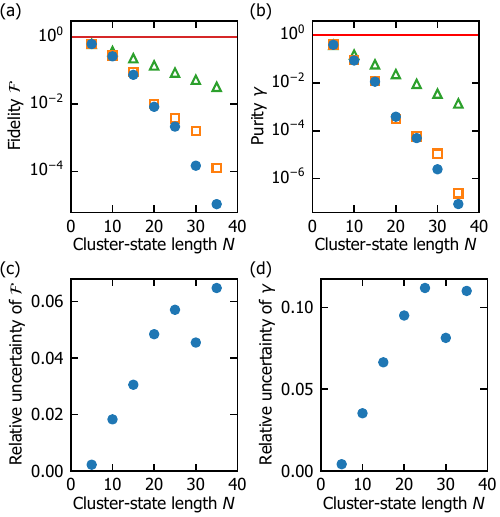}
\caption{Quantum state fidelities and purities of the generated $N$-qubit cluster states.
(a)~Fidelities $\mc{F}$ of the generated cluster states to the ideal cluster states~(blue circle).
Also plotted are the fidelities calculated using numerical models with only photon loss and dephasing errors, where the error probabilities are individually fit to each qubit for each $N$~(orange square) or are extrapolated from the average values for $N = 5$~(green triangle).
(b)~Same plot for the purity $\gamma$.
(c)~Relative uncertainty of $\mc{F}$ calculated by propagating the uncertainties of the quadrature measurements.
(d)~Same plot for the purity $\gamma$.}
\label{fig:figure5}
\end{figure}

Figures~\ref{fig:figure4}(a) and~(b) show the reconstructed density matrix of the $N$-qubit cluster state for $N = 10$ and 35.
Since the density matrices are very large~($2^N \times 2^N$), only the corners of the matrices are shown.
The complex arguments of the matrix elements show the characteristic pattern for a cluster state~(see Appendix~\ref{app:full-tomography} for the density matrix of an ideal cluster state).
The far-off-diagonal elements have small absolute values and noisy complex arguments because they are most strongly affected by any dephasing errors.
The elements in the upper-left corner have larger absolute values than the lower-right corner because of the photon loss error.

Figure~\ref{fig:figure5}(a) shows the quantum state fidelities of the $N$-qubit cluster states to the ideal cluster states.
For comparison, we also plot the fidelities calculated using numerical models which apply the photon loss and dephasing errors to each qubit in the ideal cluster state.
The error probabilities used in the numerical models are obtained from the measurements of the mean photon numbers and the stabilizer operators~(see Appendix~\ref{app:full-tomography} for details).
Figure~\ref{fig:figure5}(b) is the same plot for the purity $\gamma \coloneqq \tr[\hat{\rho}^2]$, which takes the maximum of $\gamma = 1$ for any pure state and the minimum of $\gamma = 2^{-N}$ for the completely mixed state of $N$ qubits.
For $N > 10$, the fidelities of the cluster states are significantly smaller than that of the numerical model with 9.8\% photon loss and 4.6\% phase flip errors, which are the average error probabilities obtained for $N = 5$.
By allowing the error probabilities in the numerical model to differ for each qubit and for each $N$, the fidelities and purities of the cluster states can be closely explained by the photon loss and dephasing error models for $N = 15$, 20, and 25.
This suggests that the generated cluster states for $N = 15$, 20, and 25 are affected by significantly larger incoherent errors than for $N = 5$ and 10, possibly due to heating or quasiparticle generation by the strong pulses driving the $\ket{f0}$\textrightarrow$\ket{g1}$ transition.
However, the fidelities of the 30- and 35-qubit cluster states still deviate significantly from the numerical model, which means that there are coherent error processes in addition to the increased incoherent errors.
This suggests that the gate calibrations have drifted significantly between the experiments for $N = 25$ and 30.

Figures~\ref{fig:figure5}(c) and~(d) show the relative uncertainties of the fidelity and the purity, respectively, calculated by propagating the uncertainties of the quadrature measurements.
For the fixed measurement repetition of $10^7$ per measurement setting, the relative uncertainties increase only linearly with the length of the cluster state.
This demonstrates that the tomography method is efficient in terms of the statistical uncertainty because the number of measurement repetitions required to achieve a given error tolerance can be expected to only increase quadratically with the number of qubits.

Note that the maximum length of the cluster states in this work is not limited by the computational cost of the tomography or the scaling of the statistical uncertainty but by the memory size of the arbitrary waveform generators which generated the pulse sequence.

\section{LOCALIZABLE ENTANGLEMENT}

Because the many-body density matrices of the cluster states have been reconstructed, it is possible to evaluate any entanglement criteria.
Here, we calculate the localizable entanglement, which can measure how far the entanglement persists in a chain of entangled qubits~\cite{verstraete200401entanglement,popp200504localizable}.
It is defined as the maximum entanglement that can result between two specified qubits after performing local projective measurements on the others.
The definition can vary depending on what metric is used to evaluate the resulting two-qubit entanglement.
The persistence of localizable entanglement is an important figure of merit for universal measurement-based quantum computation
using a cluster state~\cite{vandennest200610universal}.

\subsection{Localizable entanglement of the reconstructed cluster states}

\begin{figure}
\centering
\includegraphics{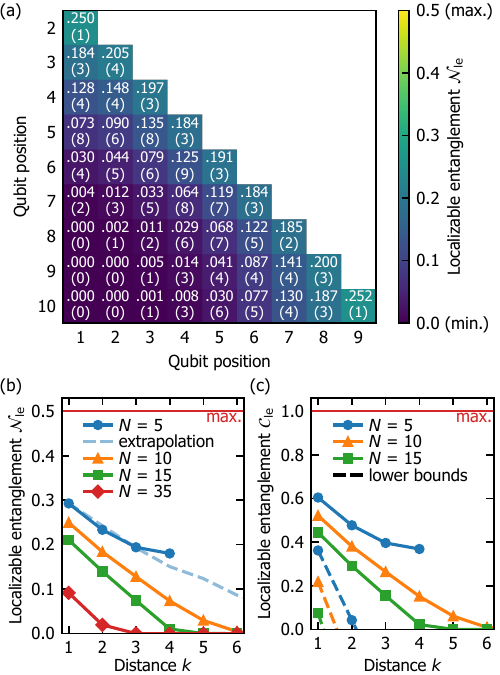}
\caption{Localizable entanglement in the generated cluster states.
(a)~Localizable entanglement in negativity between each pair of qubits in the 10-qubit cluster state.
The number in parentheses is the statistical uncertainty in the last digit.
(b)~Localizable entanglement in negativity between the first and $(1 + k)$th qubits in the cluster states~(symbols with a solid line).
For comparison, the localizable entanglement for the numerical model of a 35-qubit cluster state using the average photon loss and dephasing error probabilities obtained for the five-qubit cluster state is also shown~(dashed line).
(c)~Localizable entanglement measured in concurrence between the first and $(1 + k)$th qubits in the cluster states~(symbols with a solid line) and its lower bound based on the stabilizer measurements~(symbols with a dashed line).}
\label{fig:figure6}
\end{figure}

Let us first calculate the localizable entanglement defined in terms of the negativity $\mc{N}_\mr{le}$, which is a measure of entanglement easily computable for a two-qubit state~\cite{vidal200202computable}.
It takes the maximum $\mc{N}_\mr{le} = 0.5$ if a Bell state can result between the two specified qubits and the minimum $\mc{N}_\mr{le} = 0$ if only separable states can result.
The procedures for efficiently calculating the localizable entanglement and its uncertainty using the MPO formalism are described in Appendix~\ref{app:le}.
Figure~\ref{fig:figure6}(a) shows the localizable entanglement calculated for each pair of photonic qubits in the 10-qubit cluster state.
The localizable entanglement decays with the distance between the two qubits but persists for up to seven consecutive qubits.
For example, the localizable entanglement between the fourth and tenth qubits is $0.008 \pm 0.003$, which is significantly larger than zero.
Remarkably, even though the local correlations of only up to five consecutive qubits have been measured, the tomography method can verify the persistence of entanglement for a longer chain of qubits.

Calculating the localizable entanglement of an $N$-qubit state is computationally costly for a large $N$ because it involves a sum of $2^{N-2}$ terms corresponding to the possible outcomes of the local projective measurements.
Therefore, for $N > 15$, we estimate the localizable entanglement by randomly selecting $2^{13}$ terms and multiplying their sum by $2^{N-15}$.
Figure~\ref{fig:figure6}(b) shows the localizable entanglement between the first and $(1 + k)$th qubits in the $N$-qubit cluster state with $N = 5$, 10, 15, and 35.
We observe a decrease in the localizable entanglement for the larger cluster states.
For comparison, Fig.~\ref{fig:figure6}(b) also shows the localizable entanglement for the numerical model of a 35-qubit cluster state using the average photon loss and dephasing error probabilities obtained for the five-qubit cluster state.
The numerical model shows that, assuming uniform photon loss and dephasing error probabilities, the localizable entanglement between a pair of qubits separated by a fixed distance does not decrease with the total length of the cluster state.
The observed decrease of the localizable entanglement for larger cluster states further supports the suspicion that the coherence properties of the photon source degrade when generating a longer chain of photons.

\subsection{Lower bound based on stabilizer measurements}

Reference~\citenum{nutz201706proposal} proposed a lower bound for the localizable entanglement of a linear cluster state which can be efficiently obtained using the stabilizer measurements.
It uses the localizable entanglement defined in terms of the concurrence $\mc{C}_\mr{le}$, which is a measure of two-qubit entanglement taking the maximum of $\mc{C}_\mr{le} = 1$ for a Bell state and the minimum of $\mc{C}_\mr{le} = 0$ for a separable state~\cite{hill199706entanglement,wootters199803entanglement}.
The lower bound for the localizable entanglement between the $s$th and $(s + k)$th qubits is given by
\begin{equation}
    \mc{C}_\mr{le} \ge 1 - (k + 1)(1 - \min_r \expect{\hat{S}_r}),
\end{equation}
where
\begin{subequations} \begin{align}
    \hat{S}_1 & \coloneqq \hX_1 \hZ_2, \label{eq:stabilizers-a} \\
    \hat{S}_s & \coloneqq \hZ_{s-1} \hX_s \hZ_{s+1} \qquad (s = 2, \ldots, N-1), \\
    \hat{S}_N & \coloneqq \hZ_{N-1} \hX_N \label{eq:stabilizers-c}
\end{align} \end{subequations}
are the stabilizer operators of the $N$-qubit linear cluster state.
This lower bound does not assume that the state has an efficient MPO representation.

Figure~\ref{fig:figure6}(c) shows the localizable entanglement $\mc{C}_\mr{le}$ between the first and $(1 + k)$th qubits calculated from the reconstructed MPOs and its lower bound based on the stabilizer measurements.
The lower bound significantly underestimates the actual localizable entanglement, suggesting that it is not effective for evaluating these cluster states.

\section{DISCUSSIONS} \label{sec:discussions}

We have proposed an efficient tomography method for a sequentially generated entangled state and experimentally demonstrated it on microwave photonic cluster states.
This method is directly applicable to any physical realization of qubits which can be measured in the quadrature or Pauli basis, including optical photonic qubits in the Fock, polarization, or dual-rail encoding~\cite{kok200701linear}.
It can be straightforwardly generalized to sequentially generated qudits by using the generalized Gell--Mann basis instead of the Pauli basis.
It can also be generalized to continuous-variable systems by imposing a photon number cutoff or by choosing a finite set of basis states.
It will also be interesting to develop an analogous method for an entangled state generated using squeezers and passive linear optics by representing it as a Gaussian projected entangled-pair state~(GPEPS)~\cite{ohliger201010limitations,menicucci201104graphical}.
A GPEPS can efficiently parameterize a pure multimode Gaussian state analogously to a projected entangled-pair state~(PEPS)~\cite{verstraete200407renormalization,verstraete200412valencebond}, which is a generalization of an MPS to an arbitrary graph.
This will potentially enable an efficient tomography of the large continuous-variable cluster states which have been generated in the optical domain~\cite{yokoyama201312ultralargescale,asavanant201910generation,larsen201910deterministic}.

We make the assumption that the many-body density matrix of the qubit chain can be efficiently represented by an MPO, which we have justified by interpreting the sequential photon emission process as a tensor network.
This is a robust assumption for a time-domain-multiplexed photon source because creating an additional edge in the tensor network requires an unrealistic interaction between photonic qubits which are well-separated in time.
We have also demonstrated a way to experimentally verify this assumption by calculating the ranks of matrices constructed from local correlation measurements.
The ranks give the required bond dimension of the MPO representation, whose square root corresponds to the number of energy levels in the photon source which participated in the generation of entangled photons.
It is remarkable that such information can be obtained without directly measuring the state of the photon source.

We also make the assumption that the MPO representing the qubit chain can be reconstructed from the correlation measurements of $L = 5$ consecutive qubits.
This is a more subtle assumption, as we have seen by showing that, unlike the vast majority of other MPOs with a bond dimension of four, the ideal linear cluster state cannot be reconstructed with $L = 3$ but can be with $L = 5$.
Another subtlety is that GHZ-like entanglement spanning across more than $L$ qubits cannot be reconstructed from local correlation measurements alone because ignoring any of the qubits destroys the entanglement.
This may have affected the results presented in this work because an $N$-qubit GHZ state can be generated by omitting the $\pi_{ge} / 2$ pulses from the pulse sequence~\cite{besse202009realizing}, which implies that coherent errors in the $\pi_{ge} / 2$ pulses can introduce GHZ-like entanglement into the photon chain.
With only local correlation measurements, GHZ-like entanglement appears as increased dephasing errors, which means that the fidelities and localizable entanglements we have obtained are conservative values.
For applications where such a long-range entanglement is detrimental, global correlations should also be measured to detect the GHZ-like entanglement~\cite{baumgratz201312scalable}.

We have shown that the tomography method scales well also in terms of the statistical uncertainties of the obtained quantities, which we have calculated by propagating the uncertainties of the local correlation measurements.
The uncertainty propagation was made possible by using a least-squares method to fit the MPO to the local correlation data.
However, the least-squares method used in this work does not ensure that the best-fit MPO represents a positive semidefinite matrix, which is required for it to represent a valid density matrix.
In fact, the problem of deciding whether an MPO represents a positive semidefinite matrix has been shown to be NP-hard~\cite{kliesch201410matrixproduct}.
We argue that not enforcing the positivity is appropriate for this work because doing so has been shown to lead to an overestimation of entanglement~\cite{schwemmer201502systematic}.
Nevertheless, it may be possible to impose the positivity by using the locally-purified-density-operator~(LPDO) representation, which is positive by construction~\cite{verstraete200411matrix}.
In this case, the conditions on the bond dimension need to be further studied because an efficient MPO representation does not guarantee the existence of an efficient LPDO representation~\cite{cuevas201302purifications}.

Another interesting avenue for future work is to generalize the tomography method to entangled states with higher-dimensional graph structures.
For example, an efficient tomography method for a two-dimensional cluster state will be desired if the generation scheme demonstrated in Ref.~\citenum{osullivan202409deterministic} is scaled up to use an array of $M \gg 2$ photon emitters and an entangling gate between every neighboring pair.
For such a system, our tomography scheme currently does not scale polynomially because the bond dimension $D$ is determined by the collective system dimension of the $M$ photon emitters, which increases exponentially with $M$.
In such a case, it becomes necessary to efficiently parameterize the photonic state using the spatial locality of the entanglement among the emitters~\cite{noh202009efficient} in addition to the time-domain locality utilized in this work.
This is possible using the projected entangled-pair operator~(PEPO) representation, which can be seen as a generalization of an MPO to an arbitrary graph or a generalization of a PEPS to a mixed state~\cite{crosswhite200807finite}.
Note that the fact that a two-dimensional cluster state can be efficiently parameterized does not imply that measurement-based quantum computation is classically simulatable because it is still exponentially costly to obtain the result of the computation by contracting the two-dimensional tensor network.
Developing an efficient reconstruction algorithm for two-dimensional entangled states requires further studies because our present work makes use of the fact that a one-dimensional tensor network can be efficiently contracted.

\begin{acknowledgments}
We thank Yan Li for pointing out an error in Eq.~(10) of the preprint version, which rendered our analysis in Sec.~IV.D incorrect.
Section~IV.D was deleted from this version of the manuscript without compromising our main claims.
This work was supported in part by the University of Tokyo Program of Excellence in Photon Science~(XPS), the Japan Society for the Promotion of Science~(JSPS) Fellowship~(Grant No.\ \mbox{JP22J13650}), the Japan Science and Technology Agency~(JST) Exploratory Research for Advanced Technology~(ERATO) project~(Grant No.\ \mbox{JPMJER1601}), the Ministry of Education, Culture, Sports, Science and Technology~(MEXT) Quantum Leap Flagship Program~(Q-LEAP)~(Grant No.\ \mbox{JPMXS0118068682}), the JSPS Grant-in-Aid for Scientific Research~(KAKENHI)~(Grants No.\ \mbox{JP22H04937}, No.\ \mbox{23H03818}, and No.\ \mbox{22K18682}), the JST Center of Innovation for Sustainable Quantum AI (Grant No.\ \mbox{JPMJPF2221}).
T.\ O.\ acknowledges the support from the Endowed Project for Quantum Software Research and Education, the University of Tokyo.
\end{acknowledgments}

\appendix

\section{SAMPLE AND SETUP} \label{app:sample-and-setup}

\begin{table}
\caption{Measured sample parameters.}
\label{tab:sample-parameters}
\begin{ruledtabular}
\begin{tabular}{lcr}
$\ket{g}$--$\ket{e}$ transition frequency & $\omega_{ge} / 2 \pi$ & 8.412~GHz \\
$\ket{e}$--$\ket{f}$ transition frequency & $\omega_{ef} / 2 \pi$ & 8.031~GHz \\
\hline
$\ket{g}$--$\ket{e}$ energy relaxation time & $T_1$ & 14.3$\pm$0.9~$\mu$s \\
$\ket{g}$--$\ket{e}$ Ramsey dephasing time & $T_2^*$ & 6.9$\pm$0.7~$\mu$s \\
$\ket{g}$--$\ket{e}$ Hahn-echo dephasing time & $T_2^\mr{echo}$ & 9.9$\pm$1.4~$\mu$s \\
$\ket{e}$--$\ket{f}$ energy relaxation time & $T_{1f}$ & 15.1$\pm$1.5~$\mu$s \\
$\ket{e}$--$\ket{f}$ Ramsey dephasing time & $T_{2ef}^*$ & 5.9$\pm$0.6~$\mu$s \\
Thermal excitation ratio & $P_e / P_g$ & 0.14 \\
\hline
Resonator frequency~(dressed) & $\omega_c / 2 \pi$ & 10.6575~GHz \\
Resonator external linewidth & $\kappa_\mr{ex} / 2 \pi$ & 18.2~MHz \\
Resonator internal linewidth & $\kappa_\mr{in} / 2 \pi$ & 0.2~MHz \\
Qubit--resonator dispersive shift & $2 \chi / 2 \pi$ & $-$7.4~MHz \\
Qubit--resonator coupling strength & $g / 2 \pi$ & 239~MHz \\
\end{tabular}
\end{ruledtabular}
\end{table}

The experimental setup used in this work is nominally identical to the one used in Ref.~\citenum{sunada202204fast} except that the JPA is replaced by a similar one with a larger bandwidth and the 8--12~GHz bandpass filter between the photon source and the JPA is replaced by a 9--11~GHz bandpass filter.
The sample for the photon source is also nominally identical except that the resonator is shorter and the gap between the resonator and the coupling pin is larger.
Table~\ref{tab:sample-parameters} shows the measured sample parameters.
The frequency, external and internal linewidths, and dispersive shift of the resonator are determined using the post-selected resonator spectroscopy described in Ref.~\citenum{sunada202204fast}, which uses quantum nondemolition readouts to ensure that the qubit stays in either $\ket{g}$ or $\ket{e}$ while the reflection spectrum of the resonator is being measured.

\section{MATRIX-PRODUCT-OPERATOR FORMALISM} \label{app:mpo-formalism}

Here, we introduce the formalism of MPOs, which can efficiently represent a certain class of many-body entangled states~\cite{cirac202112matrix}.
The notation used in this work is inspired by Refs.~\citenum{baumgratz201307scalable} and \citenum{baumgratz201407efficient}.

\subsection{Matrix product state}

The MPO formalism is a generalization of the MPS formalism, which we introduce here.

A pure $N$-qubit state can be represented by a $2 \times \cdots \times 2$ complex tensor $\psi_{i_1, \ldots, i_N}$ as
\begin{equation}
    \ket{\psi} = \sum_{i_1, \ldots, i_N \in \{0, 1\}}
        \psi_{i_1, \ldots, i_N} \ket{i_1, \ldots, i_N}.
\end{equation}
An MPS representation of this state is obtained by decomposing the tensor $\psi_{i_1, \ldots, i_N}$ into a contraction of $N$ tensors: a $2 \times D$ tensor $A_1$, $D \times 2 \times D$ tensors $A_2, \cdots, A_{N-1}$, and a $D \times 2$ tensor $A_N$.
The contraction can be expressed as matrix multiplications by slicing each tensor along the two-dimensional axis:
\begin{equation}
    \psi_{i_1, \ldots, i_N} = A^{(i_1)}_1 A^{(i_2)}_2 \cdots A^{(i_{N-1})}_{N-1} A^{(i_N)}_N.
\end{equation}
Here, $A^{(i_1)}_1$ is a $D$-dimensional row vector, $A^{(i_2)}_2, \ldots, A^{(i_{N-1})}_{N-1}$ are $D \times D$ matrices, and $A^{(i_N)}_N$ is a $D$-dimensional column vector.
The dimension $D$ of the axes being contracted is called the bond dimension of the MPS.

The two-dimensional axis of the tensor $A_s$ corresponds to the basis states $\ket{0}$ and $\ket{1}$ of the $s$th qubit.
For notational convenience, let us define the ket-valued matrices
\begin{equation}
    \ms{A}_s \coloneqq \sum_{i_s \in \{0, 1\}} A^{(i_s)}_s \ket{i_s},
\end{equation}
which can be used to express the $N$-qubit ket as
\begin{subequations} \begin{align}
    \ket{\psi}
    & = \sum_{i_1, \ldots, i_N \in \{0, 1\}}
        A^{(i_1)}_1 \cdots A^{(i_N)}_N \ket{i_1, \ldots, i_N} \\
    & = \left( \sum_{i_1 \in \{0, 1\}} A^{(i_1)}_1 \ket{i_1} \right) \cdots
        \left( \sum_{i_N \in \{0, 1\}} A^{(i_N)}_N \ket{i_N} \right) \\
    & = \ms{A}_1 \cdots \ms{A}_N.
\end{align} \end{subequations}

\subsection{Matrix product operator}

Whereas an MPS can only represent a pure state, an MPO can also represent a mixed state.
If a pure state $\ket{\psi}$ can be represented by an MPS with a bond dimension of $D$, the corresponding density operator $\ketbra{\psi}{\psi}$ can be represented by an MPO with a bond dimension of $D^2$.
The definition of an MPO in the Pauli basis has been presented in Eq.~\eqref{eq:mpo}.

For notational convenience, let us define the operator-valued matrices
\begin{equation}
    \hs{A}_s \coloneqq \sum_{i_s \in \{0, 1, 2, 3\}} A^{(i_s)}_s \Pp{i_s}_s,
\end{equation}
which can be used to express the $N$-qubit density operator as
\begin{subequations} \begin{align}
    \hat{\rho} & = \frac{1}{2^N} \sum_{i_1, \ldots, i_N \in \{0, 1, 2, 3\}}
        A^{(i_1)}_1 \cdots A^{(i_N)}_N \Pp{i_1}_1 \cdots \Pp{i_N}_N \\
    & = \frac{1}{2^N} \hs{A}_1 \cdots \hs{A}_N.
\end{align} \end{subequations}

\subsection{Correlation measurement}

Given an MPO representation $A^{(i_s)}_s$ of an $N$-qubit state, the expectation value of a multi-qubit correlation measurement in the Pauli basis can be calculated as
\begin{subequations} \begin{align}
    \expect{\Pp{i_1}_1 \cdots \Pp{i_N}_N}
    & = \tr[\Pp{i_1}_1 \cdots \Pp{i_N}_N \hat{\rho}] \\
    & = A^{(i_1)}_1 \cdots A^{(i_N)}_N,
\end{align}
\end{subequations}
where we have used
\begin{equation}
    \tr[\Pp{a} \Pp{b}] = 2 \delta_{ab}
\end{equation}
for $a, b \in \{0, 1, 2, 3\}$.

\subsection{Quantum state fidelity}

\begin{figure}
\centering
\includegraphics{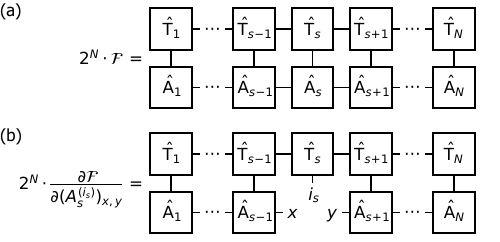}
\caption{Calculating the quantum state fidelity $\mc{F}$ of an $N$-qubit state in an MPO representation $\hs{A}_s$ to a pure target state in an MPO representation $\hs{T}_s$.
(a)~Tensor network which evaluates to $2^N \cdot \mc{F}$.
(b)~Tensor network which evaluates to the partial derivative of $2^N \cdot \mc{F}$ by a parameter of the MPO.}
\label{fig:mpo-fidelity}
\end{figure}

The quantum state fidelity of a state $\hat{\rho}$ to a pure target state $\ket{\psi}$ can be calculated as
\begin{equation}
    \mc{F} = \bra{\psi} \hat{\rho} \ket{\psi} = \tr[\ketbra{\psi}{\psi} \hat{\rho}].
\end{equation}
Given an MPO representation of the target state
\begin{equation}
    \ketbra{\psi}{\psi} = \frac{1}{2^N} \sum_{i_1, \ldots, i_N \in \{0, 1, 2, 3\}}
        T^{(i_1)}_1 \cdots T^{(i_N)}_N \Pp{i_1}_1 \cdots \Pp{i_N}_N,
\end{equation}
the quantum state fidelity can be calculated as
\begin{equation}
    \mc{F}
    = \frac{1}{2^N} \sum_{i_1, \ldots, i_N \in \{0, 1, 2, 3\}}
        T^{(i_1)}_1 \cdots T^{(i_N)}_N A^{(i_1)}_1 \cdots A^{(i_N)}_N.
\end{equation}
Figure~\ref{fig:mpo-fidelity}(a) shows the tensor-network representation of this formula.
It can be evaluated efficiently by contracting each vertical pair of tensors first.

To propagate the uncertainties in the parameters of the MPO to the uncertainty of the calculated quantum state fidelity $\mc{F}$, one needs to evaluate the partial derivatives of $\mc{F}$ by $(A^{(i_s)}_s)_{x,y}$, which denotes the $(x, y)$ matrix element of $A^{(i_s)}_s$.
This can be diagrammatically calculated from the tensor-network representation of $\mc{F}$ as shown in Fig.~\ref{fig:mpo-fidelity}(b).

\subsection{Purity}

The purity of a state $\hat{\rho}$ is defined as
\begin{equation}
    \gamma \coloneqq \tr[\hat{\rho}^2].
\end{equation}
In the MPO formalism, it can be calculated as
\begin{equation}
    \gamma
    = \frac{1}{2^N} \sum_{i_1, \ldots, i_N \in \{0, 1, 2, 3\}}
        A^{(i_1)}_1 \cdots A^{(i_N)}_N A^{(i_1)}_1 \cdots A^{(i_N)}_N.
\end{equation}
This equation is represented by the same tensor-network diagram as Fig.~\ref{fig:mpo-fidelity}(a) but with $\hs{T}_s$ replaced by $\hs{A}_s$.
The partial derivative of the purity by a matrix element of the MPO is twice the value represented by the tensor-network diagram in Fig.~\ref{fig:mpo-fidelity}(b) with $\hs{T}_s$ replaced by $\hs{A}_s$.

\subsection{Local error} \label{app:local-error}

\begin{figure}
\centering
\includegraphics{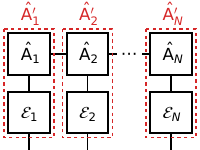}
\caption{Transformation of an MPO by local errors $\mc{E}_s$.}
\label{fig:mpo-error}
\end{figure}    

An error process on a qubit can be represented by a $4 \times 4$ real matrix $\mc{E}$, which transforms a density operator
\begin{equation}
    \hat{\rho} = \frac{1}{2} (\rho_0 \hI + \rho_1 \hX + \rho_2 \hY + \rho_3 \hZ)
\end{equation}
as
\begin{equation}
    \begin{bmatrix} \rho_0' \\ \rho_1' \\ \rho_2' \\ \rho_3' \end{bmatrix}
    = \mc{E}
    \begin{bmatrix} \rho_0 \\ \rho_1 \\ \rho_2 \\ \rho_3 \end{bmatrix}.
\end{equation}
The matrix $\mc{E}$ is called the process matrix of the error.
The unit-trace constraint $\tr[\hat{\rho}] = \rho_0 = 1$ requires that the first row of $\mc{E}$ be $[1, 0, 0, 0]$.
For the numerical models in this work, we use a combination of the amplitude damping error
\begin{equation}
    \mc{E}_\mr{ad} \coloneqq
    \begin{bmatrix}
        1 & 0 & 0 & 0 \\
        0 & \sqrt{1-\varepsilon_\mr{ad}} & 0 & 0 \\
        0 & 0 & \sqrt{1-\varepsilon_\mr{ad}} & 0 \\
        \varepsilon_\mr{ad} & 0 & 0 & 1-\varepsilon_\mr{ad}
    \end{bmatrix},
\end{equation}
where $\varepsilon_\mr{ad}$ is the probability that the qubit excitation is lost, and the pure dephasing error
\begin{equation}
    \mc{E}_\mr{pd} \coloneqq
    \begin{bmatrix}
        1 & 0 & 0 & 0 \\
        0 & 1-\varepsilon_\mr{pd} & 0 & 0 \\
        0 & 0 & 1-\varepsilon_\mr{pd} & 0 \\
        0 & 0 & 0 & 1
    \end{bmatrix},
\end{equation}
which is equivalent to the phase flip error with probability $\varepsilon_\mr{pd} / 2$.
If each qubit in an MPO experiences a single-qubit error $\mc{E}_s$, the MPO transforms as
\begin{equation}
    A'^{(i_s)}_s = \sum_{j_s \in \{0, 1, 2, 3\}} (\mc{E}_s)_{i_s,j_s} A^{(j_s)}_s.
\end{equation}
Figure~\ref{fig:mpo-error} shows the tensor-network representation of this transformation.

\section{MPO REPRESENTATION OF AN IDEAL LINEAR CLUSTER STATE} \label{app:mpo-cluster-state}

Here, we derive an MPO representation of the ideal linear cluster state, which is needed to calculate the quantum state fidelities of the photonic cluster states generated in this work.

\subsection{MPS representation}

\begin{figure}
\centering
\includegraphics{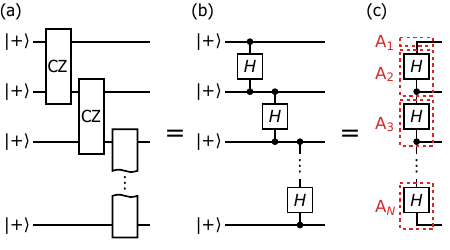}
\caption{Linear cluster state.
(a)~Quantum-circuit representation, which can also be interpreted as a tensor network.
(b)~Tensor network after applying the transformation rule shown in Fig.~\ref{fig:mps-transformations}(a).
(c)~Tensor network after applying the transformation rule shown in Fig.~\ref{fig:mps-transformations}(b).
An MPS representation $\ms{A}_s$ is obtained by contracting the Hadamard matrices $H$ with the COPY tensors.}
\label{fig:mps-cluster-state}
\end{figure}

\begin{figure}
\centering
\includegraphics{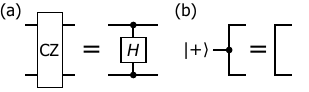}
\caption{Transformation rules used in Figs.~\ref{fig:mps-cluster-state}(a)\==(c).
(a)~A controlled-$Z$ gate equals a Hadamard matrix and two COPY tensors.
(b)~A $\ket{+}$ state and a COPY tensor equals an identity matrix.}
\label{fig:mps-transformations}
\end{figure}

We start by deriving an MPS representation of the ideal linear cluster state.
A linear cluster state belongs to a more general class of entangled states called graph states~\cite{hein200406multiparty}.
Given a graph consisting of nodes and edges, the corresponding graph state is defined as the state generated by placing a qubit in the $\ket{+} \coloneqq (\ket{0} + \ket{1})/\sqrt{2}$ state at each node and applying a controlled-$Z$~(CZ) gate wherever there is an edge between two nodes.
Note that a CZ gate is commutative and associative and therefore the order in which they are applied does not matter.
The linear cluster state is the graph state defined on a linear graph and can therefore be generated using the quantum circuit in Fig.~\ref{fig:mps-cluster-state}(a).

To find an MPS representation of the state generated by this circuit, we interpret the quantum circuit as a tensor network.
We then apply the transformation rules shown in Figs.~\ref{fig:mps-transformations}(a) and~(b) to simplify the tensor network as shown in Figs.~\ref{fig:mps-cluster-state}(b) and~(c), respectively~(see Ref.~\citenum{biamonte202001lectures} for other transformation rules).
Here, we have used the Hadamard matrix
\begin{equation}
    \raisebox{-7pt}{\includegraphics{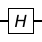}} = H
    \coloneqq \frac{1}{\sqrt{2}} \begin{bmatrix} 1 & 1 \\ 1 & -1 \end{bmatrix}
\end{equation}
and a $2 \times 2 \times 2$ tensor called the COPY tensor, which is defined in the ket-valued matrix notation as
\begin{equation}
    \raisebox{-7pt}{\includegraphics{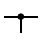}} = \ms{COPY} 
    \coloneqq \begin{bmatrix} \ket{0} & 0 \\ 0 & \ket{1} \end{bmatrix}.
\end{equation}
Note that the COPY tensor is three-fold symmetric, which can be seen by writing down its components as
\begin{equation}
    \bra{i} (\ms{COPY})_{j,k} =
    \begin{cases}
        1 & (i = j = k) \\
        0 & (\text{otherwise}).
    \end{cases}
\end{equation}

An MPS representation of the state generated by the quantum circuit is obtained by contracting the Hadamard matrix and the COPY tensor as shown in Fig.~\ref{fig:mps-cluster-state}(c):
\begin{subequations} \begin{align}
    \ms{A}_1 & = \frac{1}{\sqrt{2}} [\ket{0}, \ket{1}], \\
    \ms{A}_s & = H \cdot \ms{COPY} = \frac{1}{\sqrt{2}}
        \begin{bmatrix} \ket{0} & \ket{1} \\ \ket{0} & -\ket{1} \end{bmatrix} \quad (s = 2, \ldots, N-1), \\
    \ms{A}_N & = H \begin{bmatrix} \ket{0} \\ \ket{1} \end{bmatrix} =
        \begin{bmatrix} \ket{+} \\ \ket{-} \end{bmatrix}.
\end{align} \end{subequations}
This shows that the ideal linear cluster state is an MPS with a bond dimension of $D = 2$.

\subsection{MPO representation}

\begin{figure}
\centering
\includegraphics{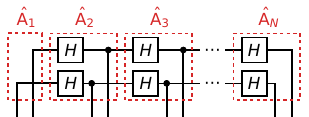}
\caption{MPO representation of a linear cluster state.} \label{fig:mpo-cluster-state}
\end{figure}

Based on the MPS representation in Fig.~\ref{fig:mps-cluster-state}(c), an MPO representation of the linear cluster state can be constructed as shown in Fig.~\ref{fig:mpo-cluster-state}.
Using the Pauli-basis representations of the Hadamard matrix, the COPY tensor, and matrix transposition
\begin{subequations} \begin{align}
    \mb{H} & \coloneqq
    \raisebox{-7pt}{\includegraphics{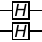}} =
    \begin{bmatrix}
        1 & 0 &  0 & 0 \\
        0 & 0 &  0 & 1 \\
        0 & 0 & -1 & 0 \\
        0 & 1 &  0 & 0
    \end{bmatrix}, \\
    \widehat{\mb{COPY}} & \coloneqq
    \raisebox{-7pt}{\includegraphics{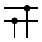}} =
    \begin{bmatrix}
        \hI &   0 &    0 & \hZ \\
          0 & \hX & -\hY &   0 \\
          0 & \hY &  \hX &   0 \\
        \hZ &   0 &    0 & \hI
    \end{bmatrix}, \\
    \mb{T} & \coloneqq
    \raisebox{-7pt}{\includegraphics{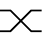}} =
    \begin{bmatrix}
        1 & 0 &  0 & 0 \\
        0 & 1 &  0 & 0 \\
        0 & 0 & -1 & 0 \\
        0 & 0 &  0 & 1
    \end{bmatrix},
\end{align} \end{subequations}
an MPO representation of the linear cluster state in the Pauli basis is obtained as
\begin{subequations} \begin{align}
    \hs{A}_1 & = [\hI_1, \hX_1, \hY_1, \hZ_1]\,\mb{T} = [\hI_1, \hX_1, -\hY_1, \hZ_1], \\
    \hs{A}_s & =
        \mb{H} \cdot \widehat{\mb{COPY}} =
        \begin{bmatrix}
            \hI_s &    0 &    0 & \hZ_s \\
            \hZ_s &    0 &    0 & \hI_s \\
              0 & -\hY_s & -\hX_s &   0 \\
              0 &  \hX_s & -\hY_s &   0
        \end{bmatrix} \notag \\
    & \qquad \qquad \qquad \qquad \qquad (s = 2, \ldots, N-1), \\
    \hs{A}_N &
        = \mb{H} \begin{bmatrix} \hI_N \\ \hX_N \\ \hY_N \\ \hZ_N \end{bmatrix}
        = \begin{bmatrix} \hI_N \\ \hZ_N \\ -\hY_N \\ \hX_N \end{bmatrix}.
\end{align} \end{subequations}
This shows that the MPO representation of an ideal cluster state has a bond dimension of $D = 4$.

\section{PROTOCOL FOR GENERATING A LINEAR CLUSTER STATE} \label{app:cluster-state-protocol}

Here, we show diagrammatically that the protocol implemented by the pulse sequence in Fig.~\ref{fig:figure2}(e) generates a linear cluster state.

\subsection{Conditional photon emission}

\begin{figure}
\centering
\includegraphics{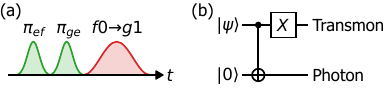}
\caption{Conditional photon emission from a transmon qubit. A photon is emitted only if the transmon starts in $\ket{e}$.
(a)~Pulse sequence. (b)~Equivalent quantum circuit.}
\label{fig:circuit-conditional-emission}
\end{figure}

The central operation of the protocol is the conditional photon emission, where the transmon qubit either emits or does not emit a photon depending on the initial qubit state.
Figure~\ref{fig:circuit-conditional-emission}(a) shows the pulse sequence for this operation, where a photon is emitted only if the transmon starts in $\ket{e}$.
If the transmon starts in a superposition of $\ket{g}$ and $\ket{e}$, this operation generates a transmon--photon entanglement as
\begin{equation}
    x \ket{g} + y \ket{e} \to x \ket{e} \otimes \ket{0}_p + y \ket{g} \otimes \ket{1}_p,
\end{equation}
where $\ket{0}_p$ and $\ket{1}_p$ denote the Fock states of the emitted photonic qubit.
Assuming that $\ket{f}$ is unpopulated at the beginning of the pulse sequence and that there is no incoming signal in the transmission line coupled to the resonator, this operation is equivalent to the quantum circuit in Fig.~\ref{fig:circuit-conditional-emission}(b).
Note that one input of the quantum circuit is fixed to $\ket{0}$, which means that this operation is a one-input two-output isometry.

\subsection{Disentangling the transmon from the photons}

\begin{figure}
\centering
\includegraphics{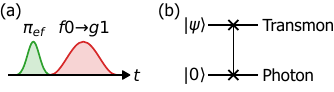}
\caption{Transferring the state of the transmon to a photonic qubit and resetting the transmon to $\ket{g}$.
(a)~Pulse sequence. (b)~Equivalent quantum circuit.}
\label{fig:circuit-1photon}
\end{figure}

At the end of the protocol, the transmon needs to be disentangled from the emitted photon chain.
This can be achieved using the pulse sequence in Fig.~\ref{fig:circuit-1photon}(a).
This pulse sequence transfers the state of the transmon to a photonic qubit and resets the transmon to $\ket{g}$ as
\begin{equation}
    x \ket{g} + y \ket{e} \to \ket{g} \otimes (x \ket{0}_p + y \ket{1}_p).
\end{equation}
Figure~\ref{fig:circuit-1photon}(b) shows an equivalent quantum circuit for this operation.

\subsection{Generating a linear cluster state}

\begin{figure}
\centering
\includegraphics{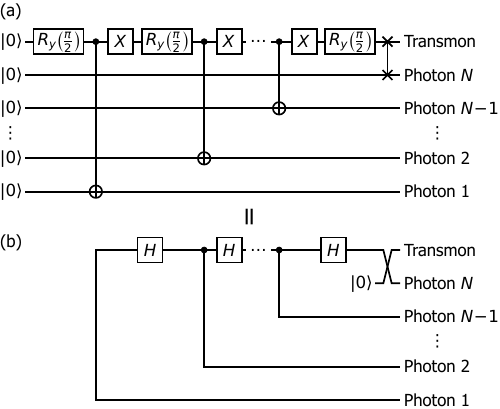}
\caption{Protocol for generating a photonic linear cluster state implemented by the pulse sequence in Fig.~\ref{fig:figure2}(e).
(a)~Quantum-circuit representation, which can also be interpreted as a tensor network.
(b)~Tensor network after applying the transformation rules shown in Figs.~\ref{fig:mps-transformations2}(a)\==(d), which matches the tensor-network representation of a linear cluster state shown in Fig.~\ref{fig:mps-cluster-state}(c).} \label{fig:mps-protocol}
\end{figure}

\begin{figure}
\centering
\includegraphics{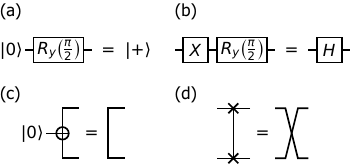}
\caption{Transformation rules used in Fig.~\ref{fig:mps-protocol}(b).
(a)~A $\ket{0}$ state and a $\pi/2$ rotation equals a $\ket{+}$ state.
(b)~An $X$ gate and a $\pi/2$ rotation equals a Hadamard gate.
(c)~A $\ket{0}$ state and an XOR tensor equals an identity matrix.
(d)~A SWAP gate equals crossed lines.}
\label{fig:mps-transformations2}
\end{figure}

Figure~\ref{fig:mps-protocol}(a) shows a quantum-circuit representation of the cluster-state generation protocol implemented by the pulse sequence in Fig.~\ref{fig:figure2}(e).
Here,
\begin{equation}
    R_y(\pi / 2) = \frac{1}{\sqrt{2}} \begin{bmatrix} 1 & -1 \\ 1 & 1 \end{bmatrix}
\end{equation}
is the quantum gate implemented by the $\pi/2$ pulse.
By applying the transformation rules shown in Figs.~\ref{fig:mps-transformations2}(a)\==(d), one can obtain the tensor-network representation of a linear cluster state shown in Fig.~\ref{fig:mps-cluster-state}(c).

\section{PAULI TOMOGRAPHY USING QUADRATURE MEASUREMENTS} \label{app:pauli-tomography}

Here, we describe how the quadrature-basis correlations shown in Eq.~\eqref{eq:quadrature-correlation} are measured and converted to the Pauli-basis correlations shown in Eq.~\eqref{eq:pauli-correlation}.

\subsection{Mode function of the photonic qubit} \label{app:mode-function}

\begin{figure}
\centering
\includegraphics{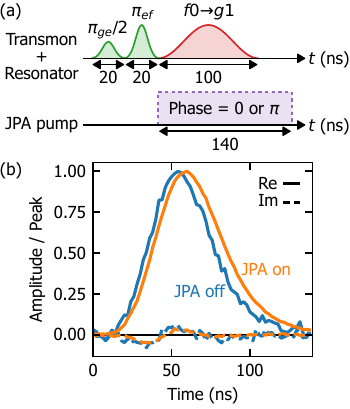}
\caption{Measuring the mode function of the photonic qubit.
(a)~Pulse sequence used to measure the averaged waveform of a photonic qubit in the $\ket{+} \coloneqq (\ket{0} + \ket{1})/\sqrt{2}$ state.
(b)~Demodulated complex waveforms measured without~(blue) and with~(orange) the amplification by the JPA.}
\label{fig:photon-waveform}
\end{figure}

First, the mode function of the photonic qubit needs to be measured because we obtain the quadrature values by amplifying the quadrature using a JPA and calculating the overlap integral between the amplified waveform and the mode function.
We use the pulse sequence shown in Fig.~\ref{fig:photon-waveform}(a) to generate a photonic qubit in the superposition state $\ket{+} \coloneqq (\ket{0} + \ket{1})/\sqrt{2}$ and measure its averaged waveform.
To obtain a complex-valued waveform, we perform another measurement with the phase of the $\pi_{ge}/2$ pulse shifted by $\pi/2$.
This generates a photonic qubit in $\ket{+i} \coloneqq (\ket{0} + i \ket{1})/\sqrt{2}$, whose averaged waveform is phase-shifted by $\pi/2$.
We also measure the waveforms of photonic qubits in $\ket{-} \coloneqq (\ket{0} - \ket{1})/\sqrt{2}$ and $\ket{-i} \coloneqq (\ket{0} - i \ket{1})/\sqrt{2}$ and subtract them from those of $\ket{+}$ and $\ket{+i}$, respectively, to remove any offset and background signal in the waveforms.
Note that the complex-valued waveform can alternatively be obtained, without phase-shifting the photon, by applying a Hilbert transform to the measured real-valued waveform, which is at the intermediate frequency (IF) of the down-converting mixer.

Figure~\ref{fig:photon-waveform}(b) shows the demodulated complex waveform obtained with and without the amplification by the JPA.
To phase-insensitively amplify the signal using the phase-sensitive mode of the JPA, we add the results of two sets of measurements with the phase of the JPA pump set to 0 and $\pi$.
This results in a phase-insensitive amplitude gain of $\sqrt{S} + 1/\sqrt{S}$, where $S$ is the squeezing factor of the JPA.
A squeezing factor of $10 \log_{10} S = 15.7$~dB is obtained from the measured waveforms.
The amplification by the JPA delays but does not significantly deform the waveform, which suggests that the gain bandwidth of the JPA is sufficient for the photonic qubit.
The waveform measured with the amplification by the JPA is used as the mode function when calculating the quadrature values in the tomography experiments.

\subsection{Pauli tomography using quadrature measurements}

To derive the conversion formula between the quadrature measurements and the Pauli measurements, let us project the quadrature operators onto the $\{ \ket{0}, \ket{1} \}$ subspace as
\begin{subequations} \begin{align}
    \hPi \hat{q}_s \hPi & = \frac{1}{\sqrt{2}} \hX_s, \label{eq:iqi-pauli-a} \\
    \hPi \hat{p}_s \hPi & = \frac{1}{\sqrt{2}} \hY_s, \\
    \hPi \frac{\hat{q}_s^2 + \hat{p}_s^2}{2} \hPi & = \hI_s - \frac{1}{2} \hZ_s, \label{eq:iqi-pauli-c}
\end{align} \end{subequations}
where $\hPi \coloneqq \ketbra{0}{0} + \ketbra{1}{1}$ is the projection operator.
By taking the expectation values of these equations, the conversion formula shown in Eqs.~\eqref{eq:pauli-quadrature-a}\==\eqref{eq:pauli-quadrature-c} is obtained.
The expectation values of the Pauli observables can therefore be estimated as
\begin{subequations} \begin{align}
    \overbar{X_s} & = \sqrt{2} \, \overbar{q_s}, \\
    \overbar{Y_s} & = \sqrt{2} \, \overbar{p_s}, \\
    \overbar{Z_s} & = 2 - \overbar{q_s^2} - \overbar{p_s^2},
\end{align} \end{subequations}
where $\overbar{q_s}$, $\overbar{p_s}$, $\overbar{q_s^2}$, and $\overbar{p_s^2}$ are the sample moments obtained by repeating the generation and quadrature measurement of the photonic qubit.
$\overbar{q_s}$ and $\overbar{q_s^2}$ are calculated from the set of measurements where the phase of the JPA pump is 0, whereas $\overbar{p_s}$ and $\overbar{p_s^2}$ are calculated from a separate set of measurements where the phase of the JPA pump is $\pi$.
Statistical uncertainties in the estimates of the Pauli observables can be calculated from the variance estimates of the sample moments as
\begin{subequations} \begin{alignat}{2}
    \se\left[\overbar{X_s}\right] & = \sqrt{\var\left[\overbar{X_s}\right]}
        && = \sqrt{2 \var\left[\overbar{q_s}\right]}, \\
    \se\left[\overbar{Y_s}\right] & = \sqrt{\var\left[\overbar{Y_s}\right]}
        && = \sqrt{2 \var\left[\overbar{p_s}\right]}, \\
    \se\left[\overbar{Z_s}\right] & = \sqrt{\var\left[\overbar{Z_s}\right]}
        && = \sqrt{\var\bigl[\overbar{q_s^2}\bigr] + \var\bigl[\overbar{p_s^2}\bigr]},
\end{alignat} \end{subequations}
where $\se$ denotes the standard error.

\subsection{Correcting for the measurement inefficiency} \label{app:eta-correction}

The previous subsection assumed that ideal quadrature measurements can be performed on a photonic qubit.
However, the photonic qubit propagates through lossy cables and noisy amplifiers before its waveform is measured.
These effects can be collectively described using the measurement efficiency $\eta$, which is equivalent to the amplitude damping error $\mc{E}_\mr{ad}$ introduced in Appendix~\ref{app:local-error} with a photon loss probability of $\varepsilon_\mr{ad} = 1 - \eta$.
Therefore, one can correct for the measurement inefficiency as
\begin{equation}
    \begin{bmatrix} 1 \\ \overbar{X_s} \\ \overbar{Y_s} \\ \overbar{Z_s} \end{bmatrix}
    = \mc{E}_\mr{ad}^{-1} \begin{bmatrix} 1 \\ \overbar{X_s'} \\ \overbar{Y_s'} \\ \overbar{Z_s'} \end{bmatrix}
    = \begin{bmatrix}
        1 \\ \eta^{-1/2} \overbar{X_s'} \\ \eta^{-1/2} \overbar{Y_s'} \\ 1 - \eta^{-1} (1 - \overbar{Z_s'})
    \end{bmatrix},
\end{equation}
where $\overbar{X_s'}$, $\overbar{Y_s'}$, and $\overbar{Z_s'}$ denote the measurement results affected by the measurement inefficiency.
Note that this correction unevenly amplifies the uncertainties in the estimates of the Pauli observables as
\begin{subequations} \begin{align}
    \se\left[\overbar{X_s}\right] & = \eta^{-1/2} \se\left[\overbar{X_s'}\right], \\
    \se\left[\overbar{Y_s}\right] & = \eta^{-1/2} \se\left[\overbar{Y_s'}\right], \\
    \se\left[\overbar{Z_s}\right] & = \eta^{-1} \se\left[\overbar{Z_s'}\right].
\end{align} \end{subequations}

\begin{figure}
\centering
\includegraphics{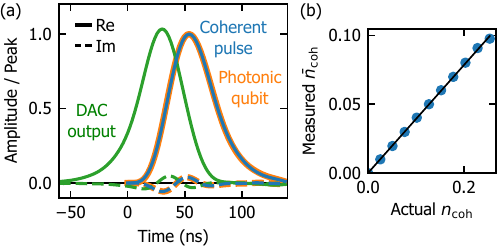}
\caption{Determining the measurement efficiency of the amplification chain using a coherent pulse.
(a)~Demodulated waveforms of a photonic qubit~(orange), a coherent pulse generated by the DAC~(green), and the coherent pulse measured using the same amplification chain as the photonic qubit~(blue).
The output waveform of the DAC is numerically pre-distorted such that the waveform of the coherent pulse matches the mode shape of the photonic qubit after it is reflected and distorted by the resonator in the photon source.
(b)~Actual and measured mean photon numbers of the coherent pulse~(blue circles) and the linear fit~(black line), whose slope gives the measurement efficiency.}
\label{fig:result-photon-undistort}
\end{figure}

The measurement efficiency $\eta$ of the amplification chain is determined by measuring a pulse mode in a coherent state with a known amplitude.
The coherent pulse is generated using the DAC and mixer which are also used to generate a readout pulse.
To accurately determine the measurement efficiency, the shape of the coherent pulse needs to match the mode shape of the photonic qubit.
However, the pulse generated by the DAC is reflected and distorted by the resonator in the photon source before reaching the amplification chain.
To compensate for this, the output waveform of the DAC is numerically pre-distorted using the inverse of the reflection coefficient $S_{11}(\omega)$ of the resonator.
Figure~\ref{fig:result-photon-undistort}(a) shows that the waveform of the coherent pulse after the reflection matches the mode shape of the photonic qubit.

The $\hat{q}$ and $\hat{p}$ quadratures of the coherent pulse are measured by setting the phase of the JPA pump to 0 and $\pi$, respectively.
After normalizing the quadrature values by equating their variances to 0.5, the mean photon number of the coherent pulse is calculated as
\begin{equation}
    \overbar{n}_\mr{coh} = \frac{1}{2} (\overbar{q}^2 + \overbar{p}^2).
\end{equation}
The measurement efficiency $\eta$ is obtained as the ratio between $\overbar{n}_\mr{coh}$ and the actual mean photon number of the coherent pulse.
The amplitude of the coherent pulse is varied to confirm that the pulse is not saturating the JPA.
Figure~\ref{fig:result-photon-undistort}(b) shows the result and the linear fit, which gives a measurement efficiency of $\eta = 0.391 \pm 0.004$.
Here, the actual mean photon number of the coherent pulse is calculated using the attenuation of the input line, which is estimated by measuring the qubit dephasing while applying a weak coherent drive at the same frequency as the pulse~\cite{kono201803quantum}.

Note that our estimation of the measurement efficiency places the reference plane at the output port of the sample.
This means that the photonic states reconstructed in this work reflect the imperfections of the photon emission process, such as the energy relaxation of the $\ket{f}$ state during the photon emission and the infidelity of the $\pi_{ef}$ pulse.

\subsection{Pauli tomography of multiple photonic qubits} \label{app:multi-pauli-tomography}

Here, we introduce the multi-qubit generalization of the conversion formula between the quadrature and Pauli measurements.
To derive the formula, let us split Eqs.~\eqref{eq:iqi-pauli-a}\==\eqref{eq:iqi-pauli-c} into two steps.
The first step converts the quadrature moments to the ``$Z$-shifted Pauli basis'' as
\begin{equation}
    \begin{bmatrix} \hI_s \\ \hX_s \\ \hY_s \\ 2\hI_s-\hZ_s \end{bmatrix} = G
    \begin{bmatrix}
        \hPi \hat{q}_s^0 \hPi \\
        \hPi \hat{p}_s^0 \hPi \\
        \hPi \hat{q}_s^1 \hPi \\
        \hPi \hat{p}_s^1 \hPi \\
        \hPi \hat{q}_s^2 \hPi \\
        \hPi \hat{p}_s^2 \hPi
    \end{bmatrix},
\end{equation}
where
\begin{equation}
    G \coloneqq
    \begin{bmatrix}
        \frac{1}{2} & \frac{1}{2} & 0 & 0 & 0 & 0 \\
        0 & 0 & \sqrt{2} & 0 & 0 & 0 \\
        0 & 0 & 0 & \sqrt{2} & 0 & 0 \\
        0 & 0 & 0 & 0 & 1 & 1
    \end{bmatrix}
\end{equation}
is the conversion matrix.
Using the notation defined in Eqs.~\eqref{eq:Q-a}\==\eqref{eq:Q-c} for the quadrature moments and the $Z$-shifted Pauli basis defined as
\begin{equation}
    \Rp{0}_s \coloneqq \hI_s, \quad
    \Rp{1}_s \coloneqq \hX_s, \quad
    \Rp{2}_s \coloneqq \hY_s, \quad
    \Rp{3}_s \coloneqq 2\hI_s-\hZ_s,
\end{equation}
the conversion can be expressed as
\begin{equation}
    \Rp{j_s}_s = \sum_{k_s \in \{0, \ldots, 5\}} G_{j_s,k_s} \hPi \Qp{k_s}_s \hPi.
\end{equation}
The second step converts the $Z$-shifted Pauli basis to the Pauli basis as
\begin{equation}
    \Pp{i_s}_s = \sum_{j_s \in \{0, 1, 2, 3\}} F_{i_s,j_s} \Rp{j_s}_s
\end{equation}
using the conversion matrix
\begin{equation}
    F \coloneqq
    \begin{bmatrix}
        1 & 0 & 0 & 0 \\
        0 & 1 & 0 & 0 \\
        0 & 0 & 1 & 0 \\
        2 & 0 & 0 & -1
    \end{bmatrix}.
\end{equation}
Using the conversion matrices $F$ and $G$, the multivariate quadrature moments can be transformed to the $Z$-shifted-Pauli-basis correlations as
\begin{subequations} \begin{multline}
    \expect{\Rp{j_1}_1 \cdots \Rp{j_N}_N} \\
    \quad = \sum_{k_1, \ldots, k_N \in \{0, \ldots, 5\}} G_{j_1, k_1} \cdots G_{j_N, k_N}
        \expect{\Qp{k_1}_1 \cdots \Qp{k_N}_N},
\end{multline} \end{subequations}
then to the Pauli-basis correlations as
\begin{subequations} \begin{multline}
    \expect{\Pp{i_1}_1 \cdots \Pp{i_N}_N} \\
    \quad = \sum_{j_1, \ldots, j_N \in \{0, 1, 2, 3\}} F_{i_1, j_1} \cdots F_{i_N, j_N}
        \expect{\Rp{j_1}_1 \cdots \Rp{j_N}_N}.
\end{multline} \end{subequations}

We have introduced the $Z$-shifted-Pauli-basis representation because it can be obtained by scaling the multivariate quadrature moments.
This means that the statistical uncertainties of the measured quadrature moments can be similarly scaled to obtain the uncertainties of the $Z$-shifted-Pauli-basis correlations.
In contrast, the conversion to the Pauli basis requires adding the zeroth and second quadrature moments, which generates correlations among the uncertainties of different Pauli observables.
By processing the data in the $Z$-shifted-Pauli-basis, we avoid the increased computational cost of propagating correlated uncertainties.

Note also that the $\Qp{i_s}_s$ basis distinguishes between $\hat{q}_s^0$ and $\hat{p}_s^0$ even though both of these equal the identity operator.
This is because, for example, $\overbar{q_1^0 q_2^1}$ and $\overbar{p_1^0 q_2^1}$ are estimated from independent experiments with different measurement settings.
The first row of the matrix $G$ takes care of averaging the results of these experiments to obtain the best estimate for $\expect{\hat{q}_2}$.
For the same reason, $\hat{q}_s^2$ and $\hat{p}_s^2$ are distinguished in the $\Qp{i_s}_s$ basis even though they are equal if the pulse mode only contains up to one photon.

\section{FULL TOMOGRAPHY OF A FIVE-QUBIT CLUSTER STATE} \label{app:full-tomography}

\begin{figure}
\centering
\includegraphics{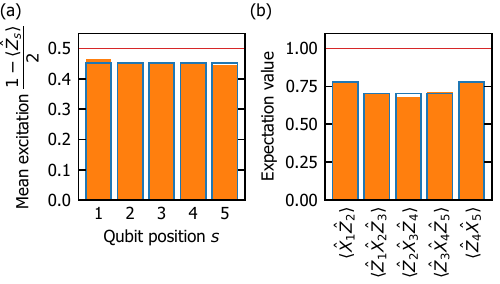}
\caption{Effects of the photon loss and dephasing errors on the generated five-qubit cluster state.
(a)~Mean excitation of each photonic qubit~(orange bar), which is smaller than the ideal value of 0.5 because of the photon loss error, and the best fit by a numerical model with uniform error probabilities across all qubits~(blue frame).
(b)~Expectation values of the stabilizer operators~(orange bar), which are smaller than the ideal value of 1 because of the photon loss and dephasing errors, and the best fit by the numerical model~(blue frame).}
\label{fig:result-5photon-ns}
\end{figure}

\begin{figure*}
\centering
\includegraphics{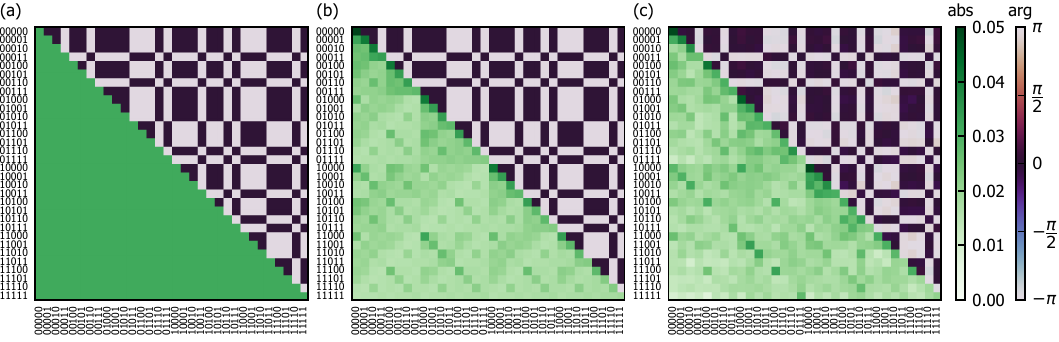}
\caption{Density matrices of (a)~an ideal five-qubit cluster state, (b)~a numerical model with uniform photon loss and dephasing errors across all qubits, and (c)~the generated cluster state.
Absolute values are plotted in the diagonal and the lower-left triangle, and the complex arguments in the upper-right triangle.}
\label{fig:result-5photon}
\end{figure*}

Here, we describe the measurements of the five-qubit linear cluster state, which can be reconstructed directly from the five-qubit correlation measurements.

We compare the reconstructed density matrix with a numerical model with uniform photon loss and dephasing probabilities across all qubits.
The photon loss error of the cluster state can be evaluated by measuring the mean excitation $(1 - \expect{\hZ_s}) / 2$ of each photonic qubit, which is affected by the photon loss but not by the dephasing error.
Figure~\ref{fig:result-5photon-ns}(a) shows the measured mean excitations and the fit, which gives an average photon loss probability of 9.8\%.
The dephasing error can be evaluated by measuring the expectation values of the stabilizer operators of the cluster state, which are affected by both the photon loss and dephasing errors.
For this, we first need to align the phase origins of the photonic qubits with the phase origins of the quadrature measurements.
This is done by numerically rotating the $s$th photonic qubit by an angle of $-\!\arctan (\expect{\hZ_{s-1} \hY_s \hZ_{s+1}} / \expect{\hZ_{s-1} \hX_s \hZ_{s+1}})$, which maximizes the stabilizer $\expect{\hZ_{s-1} \hX_s \hZ_{s+1}}$.
This correction has been applied to all data presented in this work.
Then, we obtain the stabilizers shown in Fig.~\ref{fig:result-5photon-ns}(b), which give an average dephasing error equivalent to a phase flip probability of 4.6\%.
Note that the photon loss error has a weaker effect on the first and last stabilizers because they contain fewer $\hZ$ operators than the other stabilizers.

Figures~\ref{fig:result-5photon}(a)\==(c) show the density matrices of an ideal five-qubit cluster state, the numerical model with uniform error probabilities, and the generated cluster state, respectively.
Whereas the complex arguments of the three density matrices match closely, the absolute values of the matrix elements of the experimentally generated cluster state deviate significantly from the ideal value of $2^{-5} = 0.03125$ because they are affected by the photon loss and dephasing errors.
The fidelity of the cluster state to the numerical model is $\mc{F} = 0.970$, which suggests that the numerical model captures the dominant error processes.
The fidelity of the cluster state to the ideal state is $\mc{F} = 0.616 \pm 0.006$, which is significantly larger than the threshold of 0.5 for genuine multipartite entanglement.

\section{MPO RECONSTRUCTION BY INVERSION} \label{app:mpo-inversion}

Here, we present a simplified reformulation of the MPO reconstruction procedure introduced in Ref.~\citenum{baumgratz201307scalable}.
This reformulation provides an explicit formula for the reconstructed MPO in terms of the measured local correlations, which is used in Sec.~\ref{sec:mpo-fitting} to construct an initial guess for the least-squares fitting.
It is also used to show that five-qubit local correlations are required to reconstruct a linear cluster state, despite the fact that three-qubit local correlations are sufficient for reconstructing the vast majority of MPOs with the same bond dimension.

\subsection{Using up to three-qubit local correlations}

Using the measured two- and three-qubit local correlations, let us define $4 \times 4$ matrices $B_s$ and $C^{(i)}_s$ by their elements as
\begin{subequations} \begin{align}
    (B_s)_{a,b} & \coloneqq \expect{\Pp{a}_s \Pp{b}_{s+1}} \\
    (C^{(i)}_s)_{a,b} & \coloneqq \expect{\Pp{a}_s \Pp{i}_{s+1} \Pp{b}_{s+2}}.
\end{align} \end{subequations}
Let us also define the operator-valued-matrix representation of $C^{(i)}_s$ as
\begin{equation}
    \hs{C}_s \coloneqq \sum_{i \in \{0, 1, 2, 3\}} C^{(i)}_s \Pp{i}_{s+1}.
\end{equation}
Using these matrices, an MPO representation $\hs{A}_s$ of the measured state can be reconstructed as
\begin{subequations} \begin{align}
    \hs{A}_1 & = [\hI_1, \hX_1, \hY_1, \hZ_1], \label{eq:3qubit-reconstruction-a} \\
    \hs{A}_2 & = \hs{C}_1, \\
    B_{s-1} \hs{A}_s & = \hs{C}_{s-1} \qquad (s = 3, \ldots, N-1),
        \label{eq:3qubit-reconstruction-c} \\
    \hs{A}_N & = [\hI_N, \hX_N, \hY_N, \hZ_N]^\top, \label{eq:3qubit-reconstruction-d}
\end{align} \end{subequations}
provided that the reconstructibility condition stated below is satisfied.
The reconstructibility condition guarantees that the third equation can be solved for $\hs{A}_s$.

The reconstructibility condition is expressed using an MPO representation of the true state,
\begin{subequations} \begin{align}
    \hat{\rho}
    & \coloneqq \frac{1}{2^N} \sum_{i_1, \ldots, i_N \in \{0, 1, 2, 3\}}
        T^{(i_1)}_1 \cdots T^{(i_N)}_N \Pp{i_1}_1 \cdots \Pp{i_N}_N \\
    & = \frac{1}{2^N} \hs{T}_1 \cdots \hs{T}_N,
\end{align} \end{subequations}
which always exists for a sufficiently large bond dimension $D$.
Here, one needs to choose a representation with the minimal bond dimension because the reconstructibility condition is stricter for a larger bond dimension.
Using the matrices $T^{(i_s)}_s$ of the MPO representation, let us define $4 \times D$ matrices $L_s$ by their rows as
\begin{equation}
    (L_s)_{i_s,:} \coloneqq T^{(0)}_1 \cdots T^{(0)}_{s-1} T^{(i_s)}_s
\end{equation}
and $D \times 4$ matrices $R_s$ by their columns as
\begin{equation}
    (R_s)_{:,i_s} \coloneqq T^{(i_s)}_s T^{(0)}_{s+1} \cdots T^{(0)}_N.
\end{equation}
Here, $(\cdot)_{i_s,:}$ denotes the $i_s$th row vector and $(\cdot)_{:,i_s}$ denotes the $i_s$th column vector.
Then, the reconstructibility condition is given by
\begin{subequations} \begin{alignat}{3}
    & \rank[L_s] && = D && \quad (s = 2, \ldots, N-2), \\
    & \rank[R_s] && = D && \quad (s = 3, \ldots, N-1).
\end{alignat} \end{subequations}
It immediately follows that $D \le 4$, i.e., an MPO with a bond dimension $D > 4$ cannot be reconstructed using the correlation measurements of only up to three consecutive qubits.
If the reconstructibility condition is satisfied, Eq.~\eqref{eq:3qubit-reconstruction-c} can be solved as
\begin{equation} \label{eq:3qubit-reconstruction-abc}
    \hs{A}_s = B_{s-1}^+ \hs{C}_{s-1},
\end{equation}
where ${}^+$ denotes the Moore--Penrose pseudoinverse.

\begin{figure}
\centering
\includegraphics{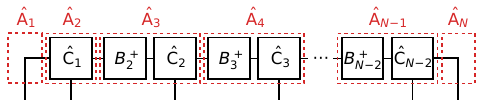}
\caption{MPO reconstructed using up to three-qubit local correlations.} \label{fig:mpo-inversion-3site}
\end{figure}

Figure~\ref{fig:mpo-inversion-3site} shows the tensor-network representation of the reconstructed MPO given by Eqs.~\eqref{eq:3qubit-reconstruction-a}\==\eqref{eq:3qubit-reconstruction-d}.
To verify this solution, let us express the measured two- and three-qubit local correlations of the true state as
\begin{subequations} \begin{align}
    B_s & = L_s R_{s+1}, \\
    \hs{C}_s & = L_s \hs{T}_{s+1} R_{s+2}.
\end{align} \end{subequations}
Because the reconstructibility condition means that $L_s$ and $R_s$ have full rank, their Moore--Penrose pseudoinverses satisfy
\begin{equation}
    L_s^+ L_s = R_s R_s^+ = I_D,
\end{equation}
where $I_D$ denotes the $D$-dimensional identity matrix, and therefore
\begin{equation}
    B_s^+ = (L_s R_{s+1})^+ = R_{s+1}^+ L_s^+.
\end{equation}
Then, one can show that Eq.~\eqref{eq:3qubit-reconstruction-abc} is a solution of Eq.~\eqref{eq:3qubit-reconstruction-c} as
\begin{subequations} \begin{align}
    B_{s-1} \hs{A}_s
    & = B_{s-1} (B_{s-1}^+ \hs{C}_{s-1}) \\
    & = (L_{s-1} R_s) (R_s^+ L_{s-1}^+) (L_{s-1} \hs{T}_s R_{s+1}) \\
    & = L_{s-1} \hs{T}_s R_{s+1} \\
    & = \hs{C}_{s-1}.
\end{align} \end{subequations}
Similarly, one can show that the reconstructed state $\hs{A}_s$ is equivalent to the true state $\hs{T}_s$ as
\begin{subequations} \begin{align}
    & \hs{A}_1 \hs{A}_2 \hs{A}_3 \cdots \hs{A}_{N-1} \hs{A}_N \\
    & = \hs{A}_1 \hs{C}_1 (B_2^+ \hs{C}_2) \cdots (B_{N-2}^+ \hs{C}_{N-2}) \hs{A}_N \\
    & = \hs{A}_1 (L_1 \hs{T}_2 R_3) (R_3^+ L_2^+) (L_2 \hs{T}_3 R_4) \cdots \notag \\
    & \qquad \qquad (R_{N-1}^+ L_{N-2}^+) (L_{N-2} \hs{T}_{N-1} R_N) \hs{A}_N \\
    & = \hs{T}_1 \hs{T}_2 \hs{T}_3 \cdots \hs{T}_{N-1} \hs{T}_N.
\end{align} \end{subequations}

Since a randomly chosen matrix almost always has full rank, an MPO with a bond dimension $D \le 4$ almost always satisfies the reconstructibility condition.
Therefore, one might expect that a linear cluster state, which has a bond dimension of $D = 4$, can be reconstructed from three-qubit local correlations.
However, it turns out that linear cluster states do not satisfy the reconstructibility condition.
This can be seen by showing that Eq.~\eqref{eq:3qubit-reconstruction-c} has no solution as follows.
Because the stabilizer operators of a linear cluster state are given by Eqs.~\eqref{eq:stabilizers-a}\==\eqref{eq:stabilizers-c}, the only nonzero two- and three-qubit local correlations of a linear cluster state are
\begin{subequations} \begin{align}
    & \expect{\hX_1 \hZ_2} = \expect{\hZ_{N-1} \hX_N} = 1, \\
    & \expect{\hY_1 \hY_2 \hZ_3} = \expect{\hZ_{N-2} \hY_{N-1} \hY_N} = 1, \\
    & \expect{\hZ_{s-1} \hX_s \hZ_{s+1}} = 1 \qquad \qquad (s = 2, \ldots, N-1).
\end{align} \end{subequations}
Therefore, the two- and three-qubit local correlation matrices are given by
\begin{equation}
    B_s =
    \begin{bmatrix}
        1 & 0 & 0 & 0 \\
        0 & 0 & 0 & 0 \\
        0 & 0 & 0 & 0 \\
        0 & 0 & 0 & 0
    \end{bmatrix}, \quad
    \hs{C}_s =
    \begin{bmatrix}
        \hI_{s+1} & 0 & 0 & 0 \\
        0 & 0 & 0 & 0 \\
        0 & 0 & 0 & 0 \\
        0 & 0 & 0 & \hX_{s+1}
    \end{bmatrix}
\end{equation}
for $s = 2, \ldots, N-3$.
Since the last column of $\hs{C}_s$ is outside the image space of $B_s$, Eq.~\eqref{eq:3qubit-reconstruction-c} has no solution.

\subsection{Using up to four-qubit local correlations}

Using the measured three- and four-qubit local correlations, let us define $4 \times 4$ matrices $B^{(i)}_1$ and $16 \times 4$ matrices $B_s$ and $C^{(i)}_s$ as
\begin{subequations} \begin{align}
    (B^{(i)}_1)_{a,b} & \coloneqq \expect{\Pp{a}_1 \Pp{i}_2 \Pp{b}_3} \\
    (B_s)_{4a+b, c} & \coloneqq \expect{\Pp{a}_s \Pp{b}_{s+1} \Pp{c}_{s+2}}, \\
    (C^{(i)}_s)_{4a+b,c} & \coloneqq \expect{\Pp{a}_s \Pp{b}_{s+1} \Pp{i}_{s+2} \Pp{c}_{s+3}}.
\end{align} \end{subequations}
Let us also define the operator-valued-matrix representations of $B^{(i)}_1$ and $C^{(i)}_s$ as
\begin{subequations} \begin{align}
    \hs{B}_1 & \coloneqq \sum_{i \in \{0, 1, 2, 3\}} B^{(i)}_1 \Pp{i}_2, \\
    \hs{C}_s & \coloneqq \sum_{i \in \{0, 1, 2, 3\}} C^{(i)}_s \Pp{i}_{s+2}.
\end{align} \end{subequations}
Using these matrices, an MPO representation $\hs{A}_s$ of the measured state can be reconstructed as
\begin{subequations} \begin{align}
    \hs{A}_1 & = [\hI_1, \hX_1, \hY_1, \hZ_1], \\
    \hs{A}_2 & = \hs{B}_1, \\
    B_{s-2} \hs{A}_s & = \hs{C}_{s-2} \qquad (s = 3, \ldots, N-1), \label{eq:4qubit-reconstruction-bac} \\
    \hs{A}_N & = [\hI_N, \hX_N, \hY_N, \hZ_N]^\top,
\end{align} \end{subequations}
provided that the reconstructibility condition stated below is satisfied.
Defining $16 \times D$ matrices $L_s$ and $D \times 4$ matrices $R_s$ as
\begin{subequations} \begin{align}
    (L_s)_{4a+b,:} & \coloneqq T^{(0)}_1 \cdots T^{(0)}_{s-1} T^{(a)}_s T^{(b)}_{s+1}, \\
    (R_s)_{:,i_s} & \coloneqq T^{(i_s)}_s T^{(0)}_{s+1} \cdots T^{(0)}_N,
\end{align} \end{subequations}
the reconstructibility condition is given by
\begin{subequations} \begin{alignat}{3}
    & \rank[L_s] && = D && \quad (s = 1, \ldots, N-3), \\
    & \rank[R_s] && = D && \quad (s = 3, \ldots, N-1).
\end{alignat} \end{subequations}
It immediately follows that $D \le 4$, i.e., an MPO with a bond dimension $D > 4$ cannot be reconstructed using the correlation measurements of only up to four consecutive qubits.
Similarly to the previous subsection, one can see that a linear cluster state does not satisfy the reconstructibility condition by showing that Eq.~\eqref{eq:4qubit-reconstruction-bac} has no solution.

\subsection{Using up to five-qubit local correlations} \label{app:mpo-inversion-5qubit}

Using the measured four- and five-qubit local correlations, let us define $4 \times 16$ matrices $B^{(i)}_1$, $16 \times 16$ matrices $B_s$ and $C^{(i)}_s$, and $16 \times 4$ matrices $B^{(i)}_{N-3}$ as
\begin{subequations} \begin{align}
    (B^{(i)}_1)_{a,4c+d} & \coloneqq \expect{\Pp{a}_1 \Pp{i}_2 \Pp{c}_3 \Pp{d}_4}, \\
    (B_s)_{4a+b, 4c+d} & \coloneqq
        \expect{\Pp{a}_s \Pp{b}_{s+1} \Pp{c}_{s+2} \Pp{d}_{s+3}}, \\
    (C^{(i)}_s)_{4a+b,4c+d} & \coloneqq
        \expect{\Pp{a}_s \Pp{b}_{s+1} \Pp{i}_{s+2} \Pp{c}_{s+3} \Pp{d}_{s+4}}, \\
    (B^{(i)}_{N-3})_{4a+b,c} & \coloneqq \expect{\Pp{a}_{N-3} \Pp{b}_{N-2} \Pp{i}_{N-1} \Pp{c}_N}.
\end{align} \end{subequations}
Let us also define the operator-valued-matrix representations of $B^{(i)}_1$, $C^{(i)}_s$, and $B^{(i)}_{N-3}$ as
\begin{subequations} \begin{align}
    \hs{B}_1 & \coloneqq \sum_{i \in \{0, 1, 2, 3\}} B^{(i)}_1 \Pp{i}_2, \\
    \hs{C}_s & \coloneqq \sum_{i \in \{0, 1, 2, 3\}} C^{(i)}_s \Pp{i}_{s+2}, \\
    \hs{B}_{N-3} & \coloneqq \sum_{i \in \{0, 1, 2, 3\}} B^{(i)}_{N-3} \Pp{i}_{N-1}.
\end{align} \end{subequations}
Using these matrices, an MPO representation $\hs{A}_s$ of the measured state can be reconstructed as
\begin{subequations} \begin{align}
    \hs{A}_1 & = [\hI_1, \hX_1, \hY_1, \hZ_1], \\
    \hs{A}_2 & = \hs{B}_1, \\
    B_{s-2} \hs{A}_s & = \hs{C}_{s-2} \qquad (s = 3, \ldots, N-2), \\
    B_{N-3} \hs{A}_{N-1} & = \hs{B}_{N-3}, \\
    \hs{A}_N & = [\hI_N, \hX_N, \hY_N, \hZ_N]^\top,
\end{align} \end{subequations}
provided that the reconstructibility condition stated below is satisfied.
Defining $16 \times D$ matrices $L_s$ and $D \times 16$ matrices $R_s$ as
\begin{subequations} \begin{align}
    (L_s)_{4a+b,:} & \coloneqq T^{(0)}_1 \cdots T^{(0)}_{s-1} T^{(a)}_s T^{(b)}_{s+1}, \\
    (R_s)_{:,4c+d} & \coloneqq T^{(c)}_s T^{(d)}_{s+1} T^{(0)}_{s+2} \cdots T^{(0)}_N,
\end{align} \end{subequations}
the reconstructibility condition is given by
\begin{subequations} \begin{alignat}{3}
    & \rank[L_s] && = D && \quad (s = 1, \ldots, N-3), \\
    & \rank[R_s] && = D && \quad (s = 3, \ldots, N-1).
\end{alignat} \end{subequations}
It immediately follows that $D \le 16$, i.e., an MPO with a bond dimension $D > 16$ cannot be reconstructed using the correlation measurements of only up to five consecutive qubits.
This time, a linear cluster state does satisfy the reconstructibility condition, as can be directly verified by substituting the MPO representation of a linear cluster state given in Appendix~\ref{app:mpo-cluster-state} into the definitions of $L_s$ and $R_s$ and seeing that they all have full rank.

\section{COMPRESSING THE BOND DIMENSION} \label{app:reducing-d}

The procedure described in Appendix~\ref{app:mpo-inversion-5qubit} constructs an MPO with a bond dimension of 16 even if the true state has a smaller bond dimension.
Here, we show that one can use the compact singular value decompositions of the four-qubit local correlation matrices to compress the bond dimension of the MPO.
We use this fact in Sec.~\ref{sec:estimating-d} to estimate the bond dimension of the measured state and in Sec.~\ref{sec:mpo-fitting} to construct an initial guess for the least-squares fitting of an MPO.

\subsection{Compression procedure}

\begin{figure}
\centering
\includegraphics{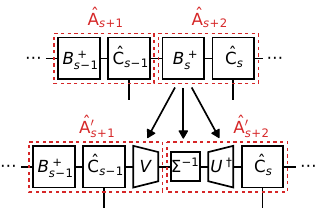}
\caption{Compressing the bond dimension between the $(s+1)$th and $(s+2)$th qubits of an MPO constructed following Appendix~\ref{app:mpo-inversion-5qubit}.} \label{fig:mpo-d-reduction}
\end{figure}

Figure~\ref{fig:mpo-d-reduction} shows the tensor-network representation of the procedure for compressing the bond dimension between the $(s+1)$th and $(s+2)$th qubits to $D_s \coloneqq \rank[B_s]$.
The compact singular value decomposition of the $16 \times 16$ matrix $B_s$ is defined as
\begin{equation}
    B_s = U \Sigma V^\dag,
\end{equation}
where $U$ and $V$ are $16 \times D_s$ semi-unitary matrices and $\Sigma$ is a $D_s \times D_s$ diagonal matrix with the positive singular values of $B_s$ on the diagonal.
Using these matrices, the Moore--Penrose pseudoinverse of $B_s$ can be expressed as
\begin{equation}
    B_s^+ = V \Sigma^{-1} U^\dag.
\end{equation}
Then, the bond dimension between the $(s+1)$th and $(s+2)$th qubits can be compressed to $D_s$ by transforming the MPO as
\begin{subequations} \begin{align}
    \hs{A}_{s+1}' & = \hs{A}_{s+1} V, \\
    \hs{A}_{s+2}' & = \Sigma^{-1} U^\dag \hs{C}_s.
\end{align} \end{subequations}

Alternatively, one can obtain an MPO with a smaller bond dimension $D'_s < \rank[B_s]$ by using a truncated singular value decomposition
\begin{equation}
    B_s \approx U \Sigma V^\dag,
\end{equation}
where $U$ and $V$ are $16 \times D'_s$ semi-unitary matrices and $\Sigma$ is a $D'_s \times D'_s$ diagonal matrix with the $D'_s$ largest singular values on the diagonal.
This gives us an approximation of the original MPO with the bond dimension between the $(s+1)$th and $(s+2)$th qubit reduced to $D'_s$.

\subsection{Estimating the bond dimension}

It follows from the above procedure that the bond dimension between the $(s+1)$th and $(s+2)$th qubits can be experimentally determined by measuring the four-qubit correlation matrix $B_s$ and counting the number of positive singular values.
However, because the measured four-qubit correlation matrix contains statistical noise, its singular values need to be significantly larger than their statistical uncertainties to qualify as being positive.
The statistical uncertainty of a measured four-qubit correlation matrix $\overbar{B}$ can be propagated to the singular values to the first order by calculating the partial derivative of a singular value $\sigma_n$ by the $(i, j)$ matrix element as
\begin{equation}
    \frac{\partial \sigma_n}{\partial \overbar{B}_{i,j}} = U^*_{i,n} V_{j,n},
\end{equation}
where $U$ and $V$ are the $16 \times 16$ unitary matrices of the singular value decomposition $\overbar{B} = U \Sigma V^\dag$.

\section{STANDARD FORM OF AN MPO} \label{app:mpo-standard-form}

When performing the least-squares fitting of an MPO, one needs to impose the constraint that the density operator represented by the MPO has unit trace.
This can be achieved by restricting the MPO to the standard form introduced in this Appendix.

\subsection{The standard form}

The standard form is given by
\begin{subequations} \begin{align}
    A^{(0)}_1 & = [1, *, \ldots, *], \\
    A^{(0)}_s & =
    \begin{bmatrix}
        1 & * & \cdots & * \\
          & * & \cdots & * \\
          &   & \ddots & \vdots \\
        0 &   &        & *
    \end{bmatrix} \quad (s = 2, \ldots, N-2), \\
    A^{(0)}_{N-1} & =
    \begin{bmatrix}
        1 & * & * & * \\
          & * & * & * \\
          &   & * & * \\
          &   &   & * \\
        0 &   &   &
    \end{bmatrix}, \\
    \hs{A}_N & = [\hI_N, \hX_N, \hY_N, \hZ_N]^\top.
\end{align} \end{subequations}
As we show later in this Appendix, any state which can be represented by an MPO with a bond dimension $D \ge 4$ has an equivalent representation in this form.
This form has fewer parameters than the original form and always satisfies the unit-trace constraint.
Furthermore, it has the useful property of
\begin{equation}
    A^{(0)}_s \cdots A^{(0)}_{N} = [1, 0, \ldots, 0]^\top,
\end{equation}
which enables one to efficiently calculate a local correlation as
\begin{equation}
    C^{(abcde)}_s = (A^{(0)}_1 \cdots A^{(0)}_{s-1}
        A^{(a)}_s A^{(b)}_{s+1} A^{(c)}_{s+2} A^{(d)}_{s+3} A^{(e)}_{s+4})_0.
\end{equation}

\subsection{Transformation into the standard form}

Given an invertible $D \times D$ matrix $U$, the transformation
\begin{subequations} \begin{align}
    \hs{A}_s' & = \hs{A}_s U, \\
    \hs{A}_{s+1}' & = U^{-1} \hs{A}_s,
\end{align} \end{subequations}
does not change the density operator represented by the MPO.
This is called a gauge transformation of the MPO.
Here, we use a series of gauge transformations to show that an MPO with a bond dimension $D \ge 4$ can be transformed into the standard form.

We first apply the transformation
\begin{subequations} \begin{align}
    (\hs{A}_{N-1}')_{:,b} & = \hs{A}_{N-1} A^{(b)}_N, \\
    \hs{A}_N' & = [\hI_N, \hX_N, \hY_N, \hZ_N]^\top.
\end{align} \end{subequations}
Next, the rectangular QR decomposition is performed on the transformed matrix $A^{(0)}_{N-1}$ as
\begin{equation}
    A^{(0)}_{N-1} = Q \begin{bmatrix} R \\ O \end{bmatrix},
\end{equation}
where $Q$ is a $D \times D$ orthogonal matrix, $R$ is a $4 \times 4$ upper triangular matrix, and $O$ is a $(D-4) \times 4$ zero matrix.
This decomposition is used to transform the MPO as
\begin{subequations} \begin{align}
    \hs{A}_{N-2}' & = \hs{A}_{N-2} Q, \\
    \hs{A}_{N-1}' & = Q^\top \hs{A}_{N-1},
\end{align} \end{subequations}
which gives us
\begin{equation}
    A^{(0)}_{N-1} = \begin{bmatrix} R \\ O \end{bmatrix}.
\end{equation}
Then, we repeat for $s = N-2, \ldots, 2$ the square QR decomposition
\begin{equation}
    A^{(0)}_s = Q R,
\end{equation}
and the transformation
\begin{subequations} \begin{align}
    \hs{A}_{s-1}' & = \hs{A}_{s-1} Q, \\
    \hs{A}_s' & = Q^\top \hs{A}_s.
\end{align} \end{subequations}
This transforms $A^{(0)}_{N-2}, \ldots, A^{(0)}_2$ into upper triangular matrices.
Finally, we use the unit-trace constraint,
\begin{subequations} \begin{align}
    1 & = \tr[\hat{\rho}] = A^{(0)}_1 \cdots A^{(0)}_N \\
    & = (A^{(0)}_1)_0 (A^{(0)}_2)_{0,0} \cdots (A^{(0)}_{N-1})_{0,0},
\end{align}
\end{subequations}
to guarantee $(A^{(0)}_1)_0, (A^{(0)}_2)_{0,0}, \ldots, (A^{(0)}_{N-1})_{0,0} \neq 0$, which enables us to rescale $\hs{A}_1, \ldots, \hs{A}_{N-1}$ as
\begin{subequations} \begin{align}
    \hs{A}_1' & = \hs{A}_1 / (A^{(0)}_1)_0, \\
    \hs{A}_s' & = \hs{A}_s / (A^{(0)}_s)_{0,0} \quad (s = 2, \ldots, N-1).
\end{align} \end{subequations}
This gives us
\begin{equation}
    (A^{(0)}_1)_0 = (A^{(0)}_2)_{0,0} = \cdots = (A^{(0)}_{N-1})_{0,0} = 1
\end{equation}
and completes the transformation into the standard form.

\section{ADDITIONAL FIGURES FOR THE 10-QUBIT CLUSTER STATE} \label{app:10-qubit}

\begin{figure}
\centering
\includegraphics{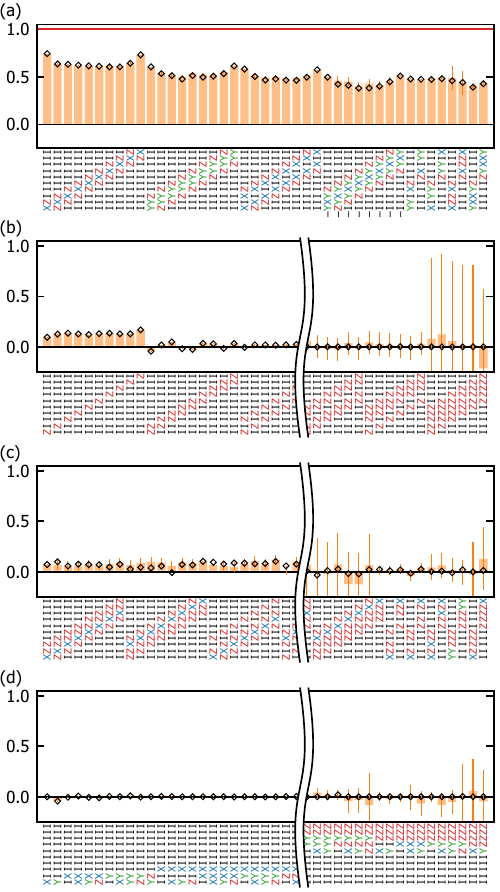}
\caption{Measured~(orange bars) and fitted~(black diamonds) local correlations for the 10-qubit cluster state.
The correction for the measurement inefficiency have been applied to the plotted values.
The error bars are calculated from the statistical uncertainties of the measured quadrature moments.
(a)~43 correlations which equal one for an ideal cluster state.
(b)~Subset of the 111 correlations which equal zero for an ideal cluster state and consist of only $\hI$ and $\hZ$ operators.
(c)~Subset of the 94 correlations which equal zero for an ideal cluster state but becomes nonzero with photon loss and dephasing errors.
(d)~Subset of the remaining 4615 correlations which equal zero for an ideal cluster state and for a cluster state affected by photon loss and dephasing errors.}
\label{fig:pauli-fit-n10}
\end{figure}

Figures~\ref{fig:pauli-fit-n10}(a)--(d) show the measured five-qubit correlations for the 10-qubit cluster state and their fits by an MPO with a bond dimension of $D = 4$.
They confirm that the least-squares fitting has successfully converged.

\begin{figure*}
\centering
\includegraphics{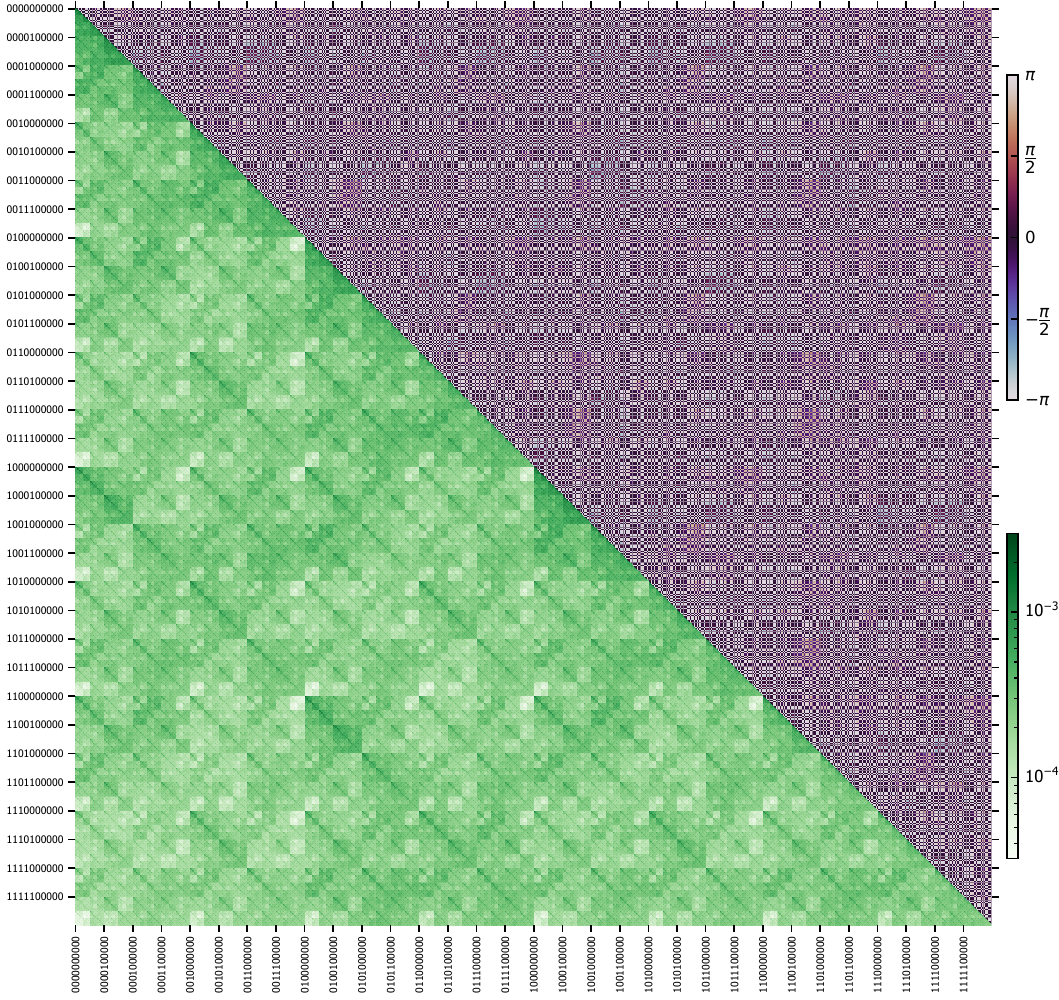}
\caption{Full density matrix of the reconstructed 10-qubit cluster state. Absolute values are plotted in the diagonal and the lower-left triangle, and the complex arguments in the upper-right triangle.}
\label{fig:rho-full-n10}
\end{figure*}

Figure~\ref{fig:rho-full-n10} shows the full density matrix of the reconstructed 10-qubit cluster state, which was partially shown in Fig.~\ref{fig:figure4}(a).

\section{LOCALIZABLE ENTANGLEMENT OF AN MPO} \label{app:le}

In this work, the metric of localizable entanglement is used to evaluate how far the entanglement persists in the generated linear cluster states.
The localizable entanglement is defined for each pair of qubits in a chain of qubits as the maximum entanglement that can be ``localized'' between the two qubits by performing local projective measurements on the other qubits~\cite{verstraete200401entanglement,popp200504localizable}.
Here, we explain how the localizable entanglement and its uncertainty can be calculated in the MPO representation.

\subsection{Local projective measurement}

\begin{figure*}
\centering
\includegraphics{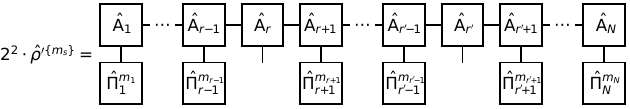}
\caption{Tensor-network representation of the unnormalized post-measurement state $\hat{\rho}'^{\{m_s\}}$, where ${\{m_s\}}$ denotes the set of measurement outcomes $m_s \in \{+1, -1\}$ obtained by performing a local projective measurement $\hat{\Pi}_s^{\pm 1}$ on every qubit of an MPO except the $r$th and $r'$th qubits.}
\label{fig:mpo-post-measurement}
\end{figure*}

A projective measurement of a qubit is defined by a pair of orthogonal projection operators $\hat{\Pi}^{\pm 1}$, where $\pm 1$ are the outcomes of the measurement.
If this measurement is performed on the $s$th qubit in an $N$-qubit state $\hat{\rho}$, the probability of obtaining each outcome is
\begin{equation}
    P(\pm 1) = \tr[\hat{\Pi}_s^{\pm 1} \hat{\rho}],
\end{equation}
and the corresponding post-measurement state of the unmeasured qubits is
\begin{equation}
    \hat{\rho}^{\pm 1} = \frac{\tr_s[\hat{\Pi}_s^{\pm 1} \hat{\rho}]}{P(\pm 1)},
\end{equation}
where $\tr_s[\cdot]$ denotes the partial trace over the $s$th qubit.
Alternatively, the unnormalized post-measurement state
\begin{equation}
    \hat{\rho}'^{\pm 1} \coloneqq \tr_s[\hat{\Pi}_s^{\pm 1} \hat{\rho}],
\end{equation}
can be used to express the outcome probability as
\begin{equation}
    P(\pm 1) = \tr[\hat{\rho}'^{\pm 1}]
\end{equation}
and the normalized post-measurement state as
\begin{equation}
    \hat{\rho}^{\pm 1} = \frac{\hat{\rho}'^{\pm 1}}{\tr[\hat{\rho}'^{\pm 1}]}.
\end{equation}

Now suppose that a local projective measurement $\hat{\Pi}_s^{\pm 1}$ is performed on every qubit of an MPO $\hs{A}_s$ except the $r$th and $r'$th qubits.
The unnormalized post-measurement state $\hat{\rho}'^{\{m_s\}}$, where ${\{m_s\}}$ denotes the set of measurement outcomes $m_s \in \{+1, -1\}$, can be calculated by evaluating the tensor network shown in Fig.~\ref{fig:mpo-post-measurement}.
Similarly to Fig.~\ref{fig:mpo-fidelity}(b), the partial derivative by a parameter of the MPO $(A^{(i_s)}_s)_{x,y}$ can be calculated by removing the tensor $\hs{A}_s$ from the diagram.
Using the unnormalized post-measurement state $\hat{\rho}'^{\{m_s\}}$, the outcome probability can be expressed as
\begin{equation}
    P(\{m_s\}) = \tr\left[\hat{\rho}'^{\{m_s\}}\right]
\end{equation}
and the normalized post-measurement state as
\begin{equation}
    \hat{\rho}^{\{m_s\}} = \frac{\hat{\rho}'^{\{m_s\}}}
        {\tr\left[\hat{\rho}'^{\{m_s\}}\right]}.
\end{equation}

\subsection{Negativity}

The entanglement between the two unmeasured qubits can be quantified using the negativity $\mc{N}$.
The negativity of a two-qubit state $\hat{\rho}$ is given by
\begin{equation}
    \mc{N}(\hat{\rho}) \coloneqq \frac{\| \rho\pt \|_1 - 1}{2},
\end{equation}
where $\rho\pt$ denotes the partial transpose of $\hat{\rho}$ in the $\{\ket{00}, \ket{01}, \ket{10}, \ket{11}\}$ basis,
\begin{equation}
    (\rho\pt)_{2i+j,2k+l} \coloneqq \bra{il} \hat{\rho} \ket{kj},
\end{equation}
and $\| \cdot \|_1$ denotes the trace norm defined as the sum of the singular values.

To propagate the uncertainties in $\hat{\rho}$ to the uncertainty of the calculated negativity, one needs to take the partial derivatives of $\| \rho\pt \|_1$ by the matrix elements of $\rho\pt$.
These can be calculated as
\begin{equation} \label{eq:trace-norm-derivative}
    \frac{\partial \| \rho\pt \|_1}{\partial (\rho\pt)_{i,j}}
    = (U V^\dagger)_{i,j},
\end{equation}
where $U$ and $V$ are the unitary matrices of the singular value decomposition $\rho\pt = U \Sigma V^\dag$.

\subsection{Localizable entanglement}

Using the negativity as the measure of two-qubit entanglement, we calculate the localizable entanglement as the expectation value of the negativity over all possible measurement outcomes,
\begin{equation} \label{eq:n-le}
    \mc{N}_\mr{le}(\hat{\rho}) \coloneqq \sum_{\{m_s\}} P(\{m_s\})\,\mc{N}(\hat{\rho}^{\{m_s\}}),
\end{equation}
given a measurement basis $\hat{\Pi}_s^{\pm 1}$ for every qubit except the two unmeasured ones at $s = r$ and $r'$.
Note that the original definition of the localizable entanglement is the maximum of $\mc{N}_\mr{le}(\hat{\rho})$ over all possible measurement bases $\{\hat{\Pi}_s^{\pm 1}\}_{s \neq r,r'}$, which is computationally costly to determine~\cite{verstraete200401entanglement}.
Because we do not perform the maximization, the definition used in this work is only a lower bound of the original definition.
Nevertheless, the optimal set of measurement bases for the ideal linear cluster state can be used to obtain a nearly maximal $\mc{N}_\mr{le}(\hat{\rho})$ for a state which is close to an ideal cluster state.
To calculate the localizable entanglement of the generated cluster states, we use the $\hX_s$ measurement basis for all qubits between the two unmeasured qubits and the $\hZ_s$ basis for all the other qubits.

To calculate the uncertainty of the localizable entanglement, Eq.~\eqref{eq:n-le} can be rewritten using the unnormalized post-measurement states $\hat{\rho}'^{\{m_s\}}$ as
\begin{equation}
    \mc{N}_\mr{le}(\hat{\rho}) = \sum_{\{m_s\}} \mc{N}(\hat{\rho}'^{\{m_s\}}).
\end{equation}
Then, Eq.~\eqref{eq:trace-norm-derivative} and the partial derivatives of Fig.~\ref{fig:mpo-post-measurement} can be used to propagate the uncertainties in the parameters of the MPO to the uncertainty of the localizable entanglement.
The localizable entanglement in terms of the concurrence can similarly be calculated using the unnormalized post-measurements states as
\begin{equation}
    \mc{C}_\mr{le}(\hat{\rho}) = \sum_{\{m_s\}} \mc{C}(\hat{\rho}'^{\{m_s\}}),
\end{equation}
where $\mc{C}(\hat{\rho})$ denotes the concurrence of a two-qubit state $\hat{\rho}$.


\FloatBarrier
\nocite{*}  
\bibliography{bibliography}  


\end{document}